%% file: nordfjord2.tex
\documentclass{svmult}
\usepackage{amssymb,epsfig}
\def\e{\mathrm{e}}

\def\N{{\mathbb N}}
\def\Z{{\mathbb Z}}
\def\R{{\mathbb R}}
\def\C{{\mathbb C}}

\def\B{{\mathcal B}}
\def\e{{\mathrm e}}
\def\d{{\mathrm d}}
\def\D{{\mathcal D}}

\def\H{{\mathcal H}}

\def\K{{\mathcal K}}
\def\Kc{{\mathcal K}^\C}
\def\L2r{L^2_{\R}}
\def\M{{\mathcal M}}

\def\P{{\mathcal P}}
\def\S{{\mathcal S}}

\def\V{{\mathcal V}}
\def\W{{\mathcal W}}
\def\Wf{W_{\mathrm{F}}}

\def\eff{{\cal F}}

\def\c{{\mathrm{c}}}

\begin{document}

\title*{Local states of free bose fields}
\author{Stephan De Bi\`evre}
\institute{UFR de Math\'ematiques et UMR P. Painlev\'e\\
Universit\'e des Sciences et Technologies de Lille\\ 59655 Villeneuve d'Ascq
Cedex France\\e-mail: Stephan.De-Bievre@math.univ-lille1.fr }

\maketitle

\section{Introduction}\label{sec:intro}
These notes contain an extended version of lectures given at the
``Summer School on Large Coulomb Systems''  in Nordfjordeid,
Norway, in august 2003. They furnish a short introduction to
some of the most basic aspects of the theory of quantum systems
that have a dynamics generated by an equation of the form
$$
\ddot q=-\Omega^2 q
$$
where $\Omega$ is a self-adjoint, positive, invertible
operator on a dense domain $\mathcal D (\Omega)$ in a real
Hilbert space $\mathcal K$.

Such systems are usually referred to as free bose fields. They are
really just harmonic systems and I will occasionally use the term
oscillator fields since I will also discuss their \emph{classical}
counterparts and because I want to stress the instructive analogy
with finite systems of coupled oscillators, which is very helpful
when one tries to understand the underlying physical
interpretation of the theory.

Many of the simplest systems of classical and quantum mechanics obey an
equation of this form. Examples include (see Sect. \ref{s:examples2}):

(i) Finite dimensional systems of coupled oscillators, where
$\K=\R^n$ and $\Omega$ is a positive definite matrix.

(ii) Lattices or chains of coupled oscillators, where $\K=\ell^2(\Z^d, \R)$
and $\Omega$ is usually a bounded operator with a possibly unbounded inverse. Those are used to model
lattice vibrations in solid state physics.

(iii) The wave equation, where $\K= L^2(K, \R)$, $K\subset\R^d$ and
$\Omega^2=-\Delta$ with suitable boundary conditions.

(iv) The massive or massless  Klein-Gordon equation on static spacetimes.
 These are a popular paradigm for studying quantum
field theory on curved spacetimes.

Despite their supposed simplicity, these systems are  interesting
for at least two reasons. First, they provide examples where the
basic concepts and methods of quantum field theory can be
explained, understood and tested. Second, they provide the
building blocks for the study of more complicated systems in
quantum field theory and (non-equilibrium) statistical mechanics,
where one or more such fields are (nonlinearly) coupled to each
other or to other, possibly finite dimensional systems. Bose
fields are for example a popular tool for modelling heath baths.
The much studied spin-bose model and more generally the
Pauli-Fierz models are all of this type.

In Section \ref{ch:classicalfree}, I shall first briefly describe the
classical mechanics of such systems in a unified way. This will then allow us
in Section \ref{ch:quantumfree} to write down the corresponding quantum
mechanical systems -- the free bose fields -- in a straightforward manner, for
{\em both infinite and finite dimensional} systems. In particular, if you are
familiar with the quantum mechanical description of finite dimensional
systems, you should conclude after reading these two sections that the
description of the infinite dimensional systems can be done quite analogously.

At that point, we will be ready to start studying the systems
constructed, and to analyze their physical properties. The only
issue I will address here, in Section \ref{ch:locobssta}, is not
one that features prominently in quantum field theory books, but
it has generated a fair amount of debate and even controversy. It
is the one of local observables, and of local states, essential
for the physical interpretation of the theory. Other topics will
be discussed in \cite{db2}. I will adopt the definition of Knight
of ``strictly local excitation of the vacuum'' (Definition
\ref{def:locexc}), that I will refer to as a strictly local or a
strictly localized state for brevity. I will then state and prove
a generalization of Knight's Theorem \cite{kn} (Sect.
\ref{s:knitherev}) which asserts that finite particle states
cannot be perfectly localized. It will furthermore be  explained
how Knight's a priori counterintuitive result can be readily
understood if one remembers the analogy between finite and
infinite dimensional harmonic systems alluded to above. I will
also discuss the link between the above result and the so-called
Newton-Wigner position operator thereby illuminating, I believe,
the difficulties associated with the latter (Sect.
\ref{s:newwig}). I will in particular argue that those
difficulties do not find their origin in special relativity or in
any form of causality violation, as is usually claimed.  It will
indeed be seen  that the Newton-Wigner position operator has an
immediate analog for a finite or infinite system of oscillators,
and that it makes absolutely no sense there since it is at odds
with basic physical intuition and since it is not compatible with
the physically reasonable definition of Knight. The conclusion I
will draw is that {\em the Newton-Wigner operator does not provide
an appropriate tool to describe the strict localization properties
of the states of extended systems of the type discussed here}. It
shows up only because of an understandable but ill-fated desire to
force too stringent a {\em particle interpretation} with all its
usual attributes on the states of a {\em field}. The right notion
of a (strictly) localized state is the one given by Knight. These
issues have generated some debate in the context of relativistic
quantum field theory over the years, upon which I shall comment in
Sect. \ref{s:newwig}.

The text is written at the graduate level and is aimed at an
audience of mathematicians with a taste  for physics and of
physicists with a taste for mathematics. A background in the
classical and quantum theory of {\em finite} dimensional systems
is assumed, although the text is basically self-contained. The
approach to the subject chosen here differs both from the usual
``second quantization'' and ``canonical quantization'' treatments
of quantum field theory prevalent in the physics literature
(although it is very close to the latter). It is not axiomatic
either. I feel it is fruitful because it allows one to apply the
intuition gained from the study of finite dimensional systems in
the infinite dimensional case. This helps in developing a good
understanding of the basic physics of quantum field theory, and in
particular to do away with some of the confusion surrounding even
some of the simplest aspects of this theory, as I hope to
illustrate with the discussion of ``localization'' in this
context. Although my approach here is resolutely non-relativistic,
I hope to show it still sheds an interesting and illuminating
light on relativistic theories as well. Indeed, the main feature
of the systems under consideration is their infinite spatial
extension, and it is this feature that distinguishes them from
systems with a finite number of particles such as atomic or
molecular systems, that have a finite spatial extension.

Related topics will be discussed in a much extended version of this manuscript, which is
in preparation \cite{db2}.

%%%%%%%%%%%%%%%%%%%%%%%%%%%%%%%%%%%%%%%%%%%%%%%%%%
%%%%%%%%%%%%%%%%%%%%%%%%%%%%%%%%%%%%%%%%%%%%%%%%%%
%%%%%%%%%%%%%%%%%%%%%%%%%%%%%%%%%%%%%%%%%%%%%%%%%%
%%%%%%%%%%%%%%%%%%%%%%%%%%%%%%%%%%%%%%%%%%%%%%%%%%

\section{Classical free harmonic systems}
\label{ch:classicalfree}
\subsection{The Hamiltonian structure}
Let us now turn to the systems described briefly in the Preface. My first goal
is to describe in detail the Hamiltonian structure underlying
\begin{equation}\label{eq:freeosc}
\ddot q +\Omega^2 q = 0.
\end{equation}
 For finite dimensional
systems, it is well known how to view (\ref{eq:freeosc}) as a Hamiltonian
system, and we will now show how to do this for infinite dimensional systems
using as only ingredient the positive operator $\Omega^2$ on
$\D(\Omega^2)\subset \K$. We need to identify a phase space on which the
solutions to this equation define a Hamiltonian flow for a suitable
Hamiltonian. For that purpose, note that, formally at least,
(\ref{eq:freeosc}) is equivalent to
$$
\dot q=p,\ \dot p =-\Omega^2 q,
$$
which are Hamilton's equations of motion for the Hamiltonian $(X=(q,p))$
\begin{equation}\label{eq:freehamiltonian}
H(X)= \frac{1}{2} p\cdot p + \frac{1}{2} q\cdot \Omega^2 q,
\end{equation}
with respect to the symplectic structure
$$
s(X,X') = q\cdot p' - q'\cdot p.
$$
Note that I use $\cdot$ for the inner product on $\K$. The Poisson
bracket of two functions $f$ and $g$ on $\K\oplus\K$ is neatly
expressed in terms of $s$ by
$$
\{f,g\} = s(\nabla_Xf, \nabla_Xg),
$$
where $\nabla_Xf = (\nabla_qf, \nabla_pf)$. Solving Hamilton's
equations of motion one obtains the Hamiltonian flow which in this
case can simply be written
\begin{equation}\label{eq:clflow}
\Phi_t = \cos \Omega t I_2 - \sin \Omega t J,
\end{equation}
where
\begin{equation}\label{eq:J}
I_2 = \left(\begin{array}{ll}1&0\\0&1
\end{array}\right)
,\quad J= \left(\begin{array}{cc}
0& -\Omega^{-1}\\
\Omega&0
\end{array}\right).
\end{equation}
For later purposes, we remark that the corresponding Hamiltonian
vector field $X_H$ defined by
\begin{equation}
\frac{\d\Phi_t}{\d t} = X_H\Phi_t.
\end{equation}
can be written
\begin{equation}\label{eq:hamfield}
X_H=-J\Omega.
\end{equation}

Of course, this is sloppy, because whereas $s$ defines a symplectic structure
on $\K \oplus\K$, the operator $J$ is not a bounded operator on $\K \oplus\K$,
so that the flow is not globally defined on this space! In other words, in the
infinite dimensional case, we have to remember that both $\Omega$ and
$\Omega^{-1}$ may be unbounded operators (think of the wave equation, for
example) and therefore we have to carefully identify a suitable phase space on
which both the symplectic structure and the flow $\Phi_t$ are globally
well-defined. For that purpose, we introduce the scale of spaces
$(\lambda\in\R)$:
$$
\K_\lambda = [\mathcal D(\Omega^\lambda)].
$$
Here the notation $[\ ]$ means that we completed $\mathcal D$ in the topology
induced by $\parallel\Omega^\lambda q\parallel$ where $\parallel \cdot
\parallel$ is the Hilbert space norm of $\K$: note that we have supposed that
$\Omega$ has a trivial kernel, so that $\parallel\Omega^\lambda
q\parallel$ defines a norm (and not just a semi-norm). Explicit
examples are developed in Sect. \ref{s:examples}.

It is easy to check that  $J$ and hence $\Phi_t$ are globally well
defined on
$$
\mathcal H = \K_{1/2} \oplus \K_{-1/2}.
$$
Moreover, the symplectic form can also defined on this space via
\begin{equation}\label{eq:symplectic}
s(X,X') =  \Omega^{1/2} q\cdot \Omega^{-1/2} p' - \Omega^{1/2} q'\cdot
\Omega^{-1/2} p.
\end{equation}
Actually, it can be checked that
$\H$ is the only space of the form $\K_\lambda \oplus \K_\mu$ with these
properties. In what follows, I shall refer to $\H$ as the (real) phase space of the
system.
Note that, from now on, whenever $w\in\K_\lambda, w'\in \K_{-\lambda}$, we
will write $w\cdot w' = \Omega^\lambda w\cdot \Omega^{-\lambda}w'$. With these
notations, one easily checks that, for $a\in\K_{1/2}, b\in\K_{-1/2}$,
\begin{equation}\label{eq:poisson}
\{b\cdot q, a\cdot p\} = a\cdot b.
\end{equation}
Here $\{\cdot, \cdot\}$ denotes the Poisson bracket.

Note that the phase space $\H$ may depend on $\Omega$, for fixed
$\K$. As long as both $\Omega$ and $\Omega^{-1}$ are bounded
operators, one has clearly $\mathcal H (\Omega) = \K \oplus \K$.
This is of course always the case when $\K$ is finite dimensional.
So for systems with a finite number of degrees of freedom, the
phase space is fixed a priori to be $\K\oplus \K$, and the
dynamics can be defined a posteriori on this fixed phase space.
However, whenever either $\Omega$ or $\Omega^{-1}$ are unbounded,
$\H(\Omega)$ differs from $\K \oplus \K$ and depends explicitly on
$\Omega$. In other words, one cannot first choose the phase space,
and then study various different dynamics on it. Instead, the
phase space and the dynamics are intimately linked: changing the
dynamics on a given fixed phase space may not make sense.

To conclude, so far, we have shown how the solutions of (\ref{eq:freeosc})
define a (linear) Hamiltonian flow $\Phi_t$ on a (real) symplectic vector
space $(\H, s)$.

As far as the classical mechanics of the system is concerned, this is really
all we need. In order to construct the corresponding quantum theory (Section
\ref{ch:quantumfree}), and in particular the quantum Hilbert space, we do
however need to exploit the structures underlying the classical theory some
more. This I will do in Sect. \ref{s:complexstructure}.
 If we were only interested in the finite dimensional case, this would
be of some interest, but not necessary. For the infinite
dimensional case it is essential. Indeed, for finite dimensional
harmonic systems, the usual Schr\" odinger quantum mechanics is of
course perfectly adequate, and the formalism developed here is
quite useless. It is however not possible to straightforwardly
adapt the Schr\" odinger formulation to the infinite dimensional
situation, and so we need to exploit the additional structures a
little  more. To understand the following developments, it is
helpful to have some examples in mind.

%%%%%%%%%%%%%%%%%%%%%%%%%%%%%%%%%%%%%%%%%%%%
%%%%%%%%%%%%%%%%%%%%%%%%%%%%%%%%%%%%%%%%%%%
\subsection{Examples}\label{s:examples}
\subsubsection{Coupled oscillators: finite dimension}\label{s:examples1}
Systems of point masses connected by springs have Hamiltonians of the type
$$
H(X) = \frac{1}{2} (p^2 + q\cdot \Omega^2 q)
$$
where $X=(q,p)\in\R^{2n}$, so that here $\K=\R^n$, and $\Omega^2$ is a
positive definite $n\times n$ matrix. More generally, this Hamiltonian arises
when linearizing any potential about a stable equilibrium point.
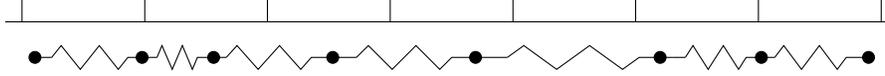
\begin{figure}
\input{chainnew.pstex_t}
\caption{A schematic representation of a chain of $8$ oscillators
moving horizontally. Linking the first to the last, you get a
ring. The tick marks indicate their equilibrium positions. In the
figure $\omega_{\mathrm w}=0$.}\label{fig:chain}
\end{figure}
An instructive example is the finite {\em oscillator chain} with periodic
boundary conditions (see Fig. \ref{fig:chain}). There, $n$ particles,
constrained to move in one dimension only,  are placed on a ring. They
interact with their nearest neighbours only, through a force that is linear in
the relative displacement of the particles and that is characterized by a
frequency $\omega_{\mathrm{n}}$. In addition they are each subjected to a
harmonic force with frequency $\omega_\mathrm{w}$.  Assuming all the particles
have identical masses, set equal to $1$, the Hamiltonian for this system reads
$$
H(X) = \frac{1}{2} \left(\sum_{i=1}^n (p(i)^2 + \omega^2_\mathrm{w} q(i)^2 +
\omega_\mathrm{n}^2(q(i+1)-q(i))^2\right).
$$
Note that in the sum the index is to be taken periodically, so that
$q(n+1)=q(1)$, {\em etc.}. I have adopted here and will continue to use the
somewhat unusual notation $v(i)$ for the $i$th component of a vector
$v\in\R^n$ or $\C^n$. This will prove very convenient later on. Introducing
$\omega_0^2=\omega_{\mathrm{w}}^2 + 2\omega_{\mathrm{n}}^2>0$ and
$$
0\leq\nu=\frac{\omega_{\mathrm{n}}^2}{\omega_0^2}\leq 1/2,
$$
the equation of motion is, for all $j=1,\dots n$,
\begin{equation}\label{eq:ring}
\ddot q(j)= -\omega_0^2 \left[q(j) -\nu(q(j+1) +
q(j-1))\right]=-(\Omega^2q)(j).
\end{equation}
One readily finds the eigenvalues of $\Omega^2$:
 they are given by
$$
\omega^2(k) = \omega_0^2[1-2\nu\cos2\pi k]\quad k=1/n, 2/n, \dots 1.
$$
Note that the eigenvalues are indeed positive, but, in order to
make sure that $0$ is not an eigenvalue, we have to impose
$\nu<1/2$, which amounts to requiring that
$\omega_{\mathrm{w}}\not=0$. This is intuitively clear: if
$\omega_{\mathrm{w}}=0$, the system allows for stationary
solutions in which all oscillators are displaced by the same
amount, so that the springs between the oscillators are not
stretched. These are are referred to as a ``zero modes''. The
above Hamiltonian provides the simplest model possible for a
harmonic crystal, and is discussed in all books on solid state
physics both from the classical and the quantum mechanical point
of view.

%%%%%%%%%%%%%%%%%%%%%%%%%%%%%%%%%%%%%%%%%%%%%%
%%%%%%%%%%%%%%%%%%%%%%%%%%%%%%%%%%%%%%%%%%%%%%%%%%%

\subsubsection{Oscillator chains and lattices}\label{s:examples2}
Having understood the finite oscillator chain, it is easy to understand the
first infinite dimensional system we shall consider, which is an infinite
linear chain of oscillators, each one linked to its neighbours and to a wall
with identical springs, so that the system is translationally invariant. The
Hamiltonian and equation of motion of this system are the same as in the case
of the ring, except that the sums now run over $\Z$. We now have
$\K=\ell^2(\Z,\R)$ and $\Omega^2$, defined precisely as in (\ref{eq:ring}) is
a bounded operator. It has a purely absolutely continuous spectrum
$\{\omega^2(k) \mid k\in [0,1]\}$, for all values of $\nu\in[0,1/2]$. Indeed,
even if $\nu=1/2$, $0$ is not an eigenvalue of $\Omega^2$, since $\eta_0$ does
not belong to $\ell^2(\Z,\R)$.

It is instructive to identify the spaces $\K_\lambda$ explicitly in this case.
For that purpose, note that the Fourier series transform
$$
\hat q(k) = \sum_{j\in\Z} q(j) \e^{-\I 2\pi jk}
$$
identifies the real Hilbert space $\ell^2(\Z,\R)$ with the real subspace of
the complex Hilbert space $L^2(\R/\Z, dk, \C)$ for which $\overline{\hat
q(k)}= \hat q(-k)$. It follows that $\K_\lambda$ can be identified with the
space of locally integrable functions $\hat q$ for which $\overline{\hat
q(k)}= \hat q(-k)$ and, more importantly, $\omega(k)^\lambda \hat q(k)$
belongs to $L^2(\R/\Z, dk, \C)$.

First of all, consider $0\leq \nu<1/2$. Then the spectrum is bounded away from
zero, which means that both $\omega(k)$ and $\omega(k)^{-1}$ are bounded
functions of $k$. As a result, then, for all $\lambda\in\R$, $\K_\lambda =
\ell^2(\Z, \R)$. In particular, then $\H =  \ell^2(\Z,\R)\times\ell^2(\Z,\R)$
and does {\em not} depend on the value of $\nu$ in the range considered.

Something interesting happens, however, if we consider the case $\nu=1/2$.
Remember that this corresponds to setting $\omega_{\mathrm{w}}=0$, which was
not allowed in the finite ring because of the existence of the zero mode. Some
remnant of this problem shows up here. Indeed, consider $\K_\lambda$, for
$\lambda<0$. Since
$$
\omega^2(k)=(2\pi\omega_0)^2k^2+\mathrm{o}(k^2),
$$
$q\in\K_\lambda$ if and only if $|k|^{\lambda}\hat q(k)$ belongs to
$L^2(\R/\Z, dk, \C)$ (and of course satisfies $\overline{\hat q(k)}= \hat
q(-k)$). But, for $\lambda=-1/2$, this is not true for all $q\in\ell^2(\Z,
\R)$. As a result, $\K=\ell^2(\Z,\R)$ is not a subspace of $\K_{-1/2}$ and
similarly $\K_{1/2}$ is not a subspace of $\K=\ell^2(\Z,\R)$. Hence the phase
space $\H$ is now different, as a set, from the phase space when $\nu\not=1/2$
and in addition, one phase space is not included in the other. To see this has
noticeable physical consequences, note the following. It seems like a
reasonable thing to wish to study the motion of the chain when initially only
one of the degrees of freedom is excited. Suppose therefore you wish to pick
the initial condition $q(i)=p(i)=0$, for all $i\not=0$, $q(0)=0\not= p(0)$. In
other words, the oscillator at the origin starts from its equilibrium position
with a non-zero initial speed, while all other oscillators are at rest at
their equilibrium positions. The trouble is that, when $\nu=1/2$, this initial
condition does not belong to the phase space! So it should be remembered that
the choice of phase space I made here, which is reasonable from many a point
of view, seems to nevertheless be somewhat too restrictive in this particular
case, since it excludes certain very reasonable initial conditions from the
state space of the system. This is one aspect of the so-called infrared problem
and it will be relevant when discussing
local observables in Section \ref{ch:locobssta}.

The generalization of the preceding considerations to $d$-dimensional
translationally invariant lattices of oscillators is immediate. One has
$\K=\ell^2(\Z^d,\R)$ and, for all $j\in\Z^d$,
\begin{equation}\label{eq:lattice}
(\Omega^2q)(j)=\omega_{\mathrm{w}}^2q(j) -
\omega_{\mathrm{n}}^2\sum_{i\in\mathrm{nn}(j)} (q(i)-q(j))=\omega_{0}^2q(j) -
\omega_{\mathrm{n}}^2\sum_{i\in\mathrm{nn}(j)}q(i),
\end{equation}
where ${\mathrm{nn}}(j)$ designates the set of nearest neighbours of $j$ and
where this time
$$
\omega_0^2 = \omega_{\mathrm{w}}^2+2d\omega_{\mathrm{n}}^2
\qquad\mathrm{and}\qquad 0\leq\nu=\frac{\omega_{\mathrm{n}}^2}{\omega_0^2}\leq
\frac{1}{2d}.
$$
Using the Fourier transform to diagonalize $\Omega^2$ one finds the dispersion
relation
$$
\omega(k)^2 = \omega_0^2\left[1-2\nu\left(\sum_{i=1}^d \cos2\pi
k_i\right)\right].
$$
This time the critical value of $\nu$ is $1/2d$ but it leads to less severe
infrared behaviour. Indeed, if $\nu=1/2d$, then
$$
\omega(k)^2 =\omega_0^2 \frac{1}{d}\sum_{i=1}^d (2\pi
k_i)^2+\mathrm{o}(|k|^2)= \frac{(2\pi\omega_0)^2}{d}|k|^2+\mathrm{o}(|k|^2).
$$
But now all compactly supported $q$ belong to $\K_\lambda$, for
all $-d/2<\lambda$, as is easily checked. As a result, this time
the phase space $\H$ contains all such initial conditions as soon
as $d\geq2$. We shall refer to them as {\em strictly local
perturbations from equilibrium} and study their quantum analogues
in Section \ref{ch:locobssta}. To be more precise, if $d\geq2$,
and if we denote by $C_{\rm c}(\Z^d)$ the space of compactly
supported sequences, then $C_{\rm c}(\Z^d) \times C_{\rm
c}(\Z^d)\subset \H$, for all possible values of $\nu$. If
$X=(q,p)\in C_{\rm c}(\Z^d)\times C_{\rm c}(\Z^d)$, then $X$
describes an initial state in which only a finite number of
oscillators is displaced from their equilibrium position and$/$or
moving. So for this rather large and very natural class of initial
conditions, the dynamics can be investigated as a function of
$\nu$, for all possible values of $\nu$.

Lattices of oscillators are used to describe the thermal and
acoustic properties of various solids, such as metals, crystals of
all sorts, amorphous materials {\em etc.}.  Putting
$\omega_\mathrm{n}=0$ in the expressions above, one obtains the
so-called Einstein model, in which the oscillators representing
the ions of the solid are not coupled. The case where
$\omega_\mathrm{n}\not=0$ is the Debye model. In more
sophisticated models still, different geometries may appear
(hexagonal lattices, body or face centered cubic lattices {\em
etc.}), and the spring constants may vary from site to site in
periodic, quasi-periodic or random ways.

%%%%%%%%%%%%%%%%%%%%%%%%%%%%%%%%%%%%%%%%%%%%%%%%%%%%%%
%%%%%%%%%%%%%%%%%%%%%%%%%%%%%%%%%%%%%%%%%%%%%%%%%%%%%%%

\subsubsection{Wave and Klein-Gordon equations}\label{s:examples3}
The wave equation
$$
\partial_t^2 q(x,t) = \Delta q(x,t)
$$
on a domain $K\subset\R^d$ with Dirichlet boundary conditions is another
example of a free oscillator field where $\K=L^2(K, \R)$ and $\D(\Omega)$ is
the domain of the square root of the Dirichlet Laplacian. When $K$ is a
bounded set, the spectrum of the Dirichlet Laplacian is  discrete. No infrared
problem then arises, reflecting the fact that no arbitrary long wavelengths
can occur in the system.

The case where $K=\R^d$ is instructive and easy to work out thanks
to its translational invariance. The situation is completely
analogous with the one in Sect. \ref{s:examples2}. Writing
$\omega(k)=\sqrt{k^2}$, the space $\K_\lambda$ is for each real
$\lambda$ naturally isomorphic to the real subspace of  $L^2(\R^d,
\omega(k)^{2\lambda}dk, \C)$ given by the condition
$\overline{\hat q(k)}=q(-k)$. If $d\geq 2$, the Schwartz space is
a subspace of $\K_{\pm\frac12}$.

One can also consider the more general case where $K$ is a Riemannian manifold
with metric $\gamma$ and $-\Delta$ the corresponding Laplace-Beltrami
operator. Replacing $-\Delta$ by $-\Delta + m^2$ $(m>0)$ in the above, one
obtains the Klein-Gordon equation. It plays an important role in the
relativistic quantum field theory on flat or curved spacetimes.

%%%%%%%%%%%%%%%%%%%%%%%%%%%%%%%%%%%%%%%%%%%%
%%%%%%%%%%%%%%%%%%%%%%%%%%%%%%%%%%%%%%%%%%%

\subsection[Complex structure on phase space]
{A preferred complex structure on the {\em real} classical phase space $\H$}
\label{s:complexstructure} The simple linear systems we are dealing with here
have some extra structure that is encoded in the matrix $J$ defined in
(\ref{eq:J}). Noticing that $J^2=-I_2$, one sees $J$ defines an $s$-compatible
({\em i.e.} $s(JX,JY)=s(X,Y)$) and positive definite ({\em i.e.} $s(X,JX)\geq
0$ and $s(X,JY)=0, \forall Y\in\H$ implies $X=0$) complex structure on $\H$.
As a result, $\H$ can first of all be viewed as a real Hilbert space, with
inner product
\begin{equation}\label{eq:gomega}
g_\Omega(X,Y)\stackrel{\mathrm{def}}{=} s(X,JY)=\sqrt\Omega q\cdot \sqrt\Omega
q' + \sqrt\Omega^{-1} p \cdot \sqrt\Omega^{-1} p',
\end{equation}
where $Y=(q', p')$. Of course, we recognize here the natural inner product on
$\H=\K_{1/2}\oplus \K_{-1/2}$, written in terms of the symplectic form and
$J$.

In addition, $J$ can be used to equip $\H$ with a {\em complex}
Hilbert space structure, where multiplication with the complex
number $a +\I b\in\C$ is defined by
$$
(a+ \I b)X \stackrel{\mathrm{def}}{=} (a+bJ)X, \quad \forall X\in
\H
$$
and with the inner product
\begin{equation}\label{eq:complexH}
\langle X, Y\rangle_+ = \frac{1}{2}(g_\Omega(X,Y) + \I s(X,Y)).
\end{equation}
 Note that, when $\H$ has $2n$
{\em real} dimensions, the complex vector space $(\H, J)$ has only $n$ complex
dimensions.

Since $\Phi_t$ is symplectic and commutes with $J$, one easily checks that
$$
g_\Omega(\Phi_t X, \Phi_t Y)=g_\Omega(X, Y)\quad \mathrm{and}\quad\langle
\Phi_t X, \Phi_t Y \rangle_+ = \langle X, Y \rangle_+,
$$
so that $\Phi_t$ is a unitary operator on the complex Hilbert
space $(\H, J, \langle \cdot, \cdot \rangle_+)$. As a result,
$X_H=-J\Omega$, the generating Hamiltonian vector field is
necessarily anti-self-adjoint and one can check that in addition
\begin{equation}\label{eq:energycond}
\I\langle X, X_H X\rangle_+ = H(X).
\end{equation}

It is natural to wonder if there exist many  complex structures on $\H$ with
these properties. In fact,  $J$ is  the unique $s$-compatible, positive
complex structure on $\H$ so that $\Phi_t$ is unitary on the corresponding
complex Hilbert space \cite{db2}. In other
words, the phase space $\H$ of an oscillator field, which is a {\em real}
symplectic space, carries a natural, flow-invariant complex Hilbert space
structure!

The ensuing complex Hilbert space seems a somewhat abstract
object, but it can be naturally identified with $\K^\C$, the
complexification of $\K$, as I now explain. In the following,
whenever $V$ is a real vector space, $V^\C=V \oplus \I V$ will
denote its complexification. In the concrete examples I have in
mind, where $V=\K=\R^n, \ell^2(\Z^d, \R)$ or $L^2(\R^d, \R)$, one
finds $V^\C=\K^\C= \C^n, \ell^2(\Z^d, \C)$ or $L^2(\R^d, \C)$,
respectively. The identification goes as follows~:
\begin{equation}\label{eq:zdef}
z_\Omega: X=(q,p)\in\H\mapsto
z_\Omega(X)=\frac{1}{\sqrt2}(\sqrt\Omega q + \I
\frac{1}{\sqrt\Omega} p) \in\K^\C.
\end{equation}
The following proposition is then easily proven.
\begin{proposition} \label{prop:hiskc} The map $z_\Omega$ defines an isomorphism between the complex Hilbert
spaces $(\H, J, \langle\cdot, \cdot\rangle_+)$ and $\Kc$,
intertwining the dynamics $\Phi_t$ with $\e^{-\I \Omega t}$. More
precisely,
\begin{equation}\label{eq:zprop1}
z_\Omega(JX) = \I z_\Omega(X)\quad \overline{z_\Omega(X)}\cdot
z_\Omega(X')= \langle X, X'\rangle_+.
\end{equation}
and
\begin{equation}\label{eq:zprop2}
z_\Omega(\Phi_tX) = \e^{-\I \Omega t} z_\Omega(X).
\end{equation}
\end{proposition}
Note that $\cdot$ has been extended to $\K^\C$ by linearity in each variable
so that the inner product on $\K^\C$ is given by $\bar z\cdot z'$, for
$z,z'\in\Kc$. The choice of the unnatural looking factor $1/\sqrt 2$  in the
definition of $z_\Omega$ and of the matching factor $1/2$ in $\langle
X,X'\rangle_+$ are conventions chosen to make comparison to the physics
literature simple, as we will see further on. Similarly, for later purposes,
we define
\begin{equation}\label{eq:zdagdef}
z^{\dagger}_\Omega: X=(q,p)\in\H\mapsto
z_\Omega^\dagger(X)=\frac{1}{\sqrt2}(\sqrt\Omega q - \I
\frac{1}{\sqrt\Omega} p)\in\K^\C,
\end{equation}
which is complex anti-linear
\begin{equation}\label{eq:zdagprop1}
z_\Omega^\dagger(JX) = -\I z_\Omega^\dagger(X)
\end{equation}
and
\begin{equation}\label{eq:zdagprop2}
\overline{z_\Omega^\dagger(X)}\cdot z_\Omega^\dagger(X')= \overline{\langle X,
X'\rangle_+}.
\end{equation}

The linear map $z_\Omega$ is readily inverted and one has, in obvious
notations
\begin{equation}\label{eq:zinvert}
q=\frac{1}{\sqrt{2\Omega}}(z_\Omega(X) + z_\Omega^\dagger(X))\
\mathrm{and}\ p=\frac{\sqrt\Omega}{\I\sqrt2}(z_\Omega(X)
-z_\Omega^\dagger(X)),
\end{equation}
and
\begin{equation}\label{eq:hamzomega}
H(X)= z_\Omega(X)^\dagger\cdot \Omega z_\Omega(X).
\end{equation}

In conclusion, we established that, having started with a {\em real} Hilbert
space $\K$ and a positive self-adjoint operator $\Omega$, the classical phase
space $\H$ of the corresponding oscillator equation $\ddot q=-\Omega^2 q$ can
be identified naturally with the {\em complex} Hilbert space $\Kc$, on which
the dynamics is simply the unitary group generated by $\Omega$,  the
symplectic structure is the imaginary part of the inner product and the
Hamiltonian is given by $H(z)=\overline z\cdot \Omega z$. We therefore ended
up with a {\em mathematically} completely
 equivalent description of the original phase space $\H$, its symplectic
structure and the dynamics $\Phi_t$ generated by the Hamiltonian $H$ in
(\ref{eq:freehamiltonian}).

It is however important to understand  that the {\em physical} interpretation
of this new formulation should be done carefully, as I  explain in Sect.
\ref{s:physint1}.

%%%%%%%%%%%%%%%%%%%%%%%%%%%%%%%%%%%%%%%%%%%%%
%%%%%%%%%%%%%%%%%%%%%%%%%%%%%%%%%%%%%%%%%%%%%%
%%%%%%%%%%%%%%%%%%%%%%%%%%%%%%%%%%%%%%%%%%%%%%%%%%%

\subsection{Physical interpretation}\label{s:physint1}
It is  instructive to first look at what the formalism of the Sect.
\ref{s:complexstructure} yields for {\em finite dimensional} systems of
coupled oscillators, such as the oscillator ring. In that case $\K=\R^n$ and
hence $\Kc=\C^n$. Note however that the identification of $\R^{2n}$ with
$\C^n$ depends in a non-trivial way on $\Omega$ which makes a direct
interpretation of points of $\C^n$ difficult. In particular, let
$X=(q,p)\in\R^{2n}=\H$. Then the components of $q$ and $p$ have a direct
physical interpretation as the displacements and momenta of the different
oscillators. The $i$th component of the corresponding vector $z=z_\Omega(X)\in
\C^n$ does not have such a direct simple interpretation
 since it is not a function
of the diplacement $q_i$ and momentum $p_i$ of the $i$th
oscillator alone, but it is a function of  the diplacements $q_j$
and momenta $p_j$ of all the oscillators.  This is so because in
general, the matrix $\Omega^{1/2}$ has no (or few) zero
off-diagonal entries, even if $\Omega^2$ is tri-diagonal, as in
the oscillator chain. Indeed, in that case, $\Omega^2$ is a
difference operator, but $\Omega^{1/2}$ is not. Conversely, as is
clear from (\ref{eq:zinvert}), $q_i$ and $p_i$ depend on all
components of $z_\Omega(X)$, not only on the $i$th one.  This
explains why the alternative formulation of the problem in terms
of the complex space $\Kc=\C^n$ is not found in classical
mechanics textbooks. Indeed, one is typically interested in
questions concerning the displacements of the different
oscillators, the energy distribution over the oscillators when the
system is in a normal mode, energy propagation along the
oscillators when originally only one oscillator is excited, {\em
etc.}. Such questions are obviously more easily addressed in the
original formulation.

Another way to see why the alternative formulation leads to interpretational
problems is as follows. Suppose we are studying two oscillator systems, one
with potential $\frac12q\cdot\Omega^2q$ and another with
$\frac12q\cdot{\Omega'}^2q$, where $\Omega^2\not={\Omega'}^2$. To fix ideas,
we can think of ${\Omega'}^2$ as being a perturbation of $\Omega^2$ which is
obtained by changing just one spring constant.
 Suppose now that the state of the first system is $z\in\Kc$, and of the second is
$z'\in\Kc$. Suppose $z=z'=z_0\in \C^n$. Would you say the two systems are in
the same state? Certainly not in general! Indeed, as a result of what
precedes, and in particular of (\ref{eq:zinvert}), the same point
$z_0\in\K^\C$ yields entirely different values for the displacements $q_i,
q_i'$ and the momenta $p_i, p_i'$ of the two oscillator systems! Indeed, we
would normally say that the two systems are in the same state if the positions
$q_i, q_i'$ and momenta $p_i, p_i'$ of the different degrees of freedom take
the same values, that is to say if $X=X'$. But that is not the same as saying
$z=z'$.  In other words, if you decide to say $\Kc$ is the phase space of your
system, you should always remember that the physical interpretation of its
points {\em depends on the dynamics, i.e.} on $\Omega$. A similar phenomenon
produces itself in the quantum mechanical description of oscillator systems as
we will see in Sect. \ref{s:focrepfin}.

Suppose now we deal with an infinite dimensional oscillator field, such as an
oscillator chain or a wave equation. As in the finite dimensional case, the
elements of $\H$ then have a direct interpretation in terms of oscillator
displacements, wave propagation {\em etc.}, whereas those of $\Kc$ don't. But
now an additional complicating phenomenon that we already pointed out occurs:
starting with a fixed $\K$, different choices of $\Omega$ may lead to
different phase spaces $\H$! We gave an example for the oscillator chain in
Sect. \ref{s:examples2}. Talking about ``the same state'' for different
systems now becomes very difficult, since the state space $\H$ depends on the
system considered.  It is then tempting to prefer the alternative formulation
where the phase space $\Kc$ is independent of the dynamics, but at that point
it should always be remembered that the same point in $\Kc$ has a different
interpretation depending on which system you consider.

In spite of those interpretational difficulties,  the alternative formulation
of the classical mechanics of oscillator systems will
 turn out to be useful (and even crucial)
in the quantum mechanical description of oscillator fields. Indeed, the
quantum Hilbert space for the free oscillator field will be seen to be the
symmetric Fock space over $(\H, J, \langle\cdot, \cdot\rangle_+)$ (see Section
\ref{ch:quantumfree}). But identifying the latter with $\Kc$ allows one to
conveniently identify the quantum Hilbert space as the symmetric Fock space
over $\Kc$. This way, one can work on a fixed Hilbert space, while changing
the dynamics by perturbing $\Omega$, for example. This is very convenient.
Still, the rather obvious, seemingly trivial and innocuous remarks above
concerning the interpretation of the {\em classical} field theory  are  at the
origin of further, more subtle interpretational difficulties with the quantum
field theory of infinite dimensional oscillator fields as well, to which I
shall come back in Sects. \ref{s:focrepfin} and \ref{s:parint}.

%%%%%%%%%%%%%%%%%%%%%%%%%%%%%%%%%%%%%%%%%%%%%%%%%%%%
%%%%%%%%%%%%%%%%%%%%%%%%%%%%%%%%%%%%%%%%%%%%%%%%%%%%

\subsection{Creation and annihilation functions on $\H$}\label{s:creationclassical}
For the purposes of quantum mechanics, it will turn out to be convenient to
develop the previous considerations
 somewhat further.
Everybody is familiar with creation and annihilation operators in quantum
mechanics. These objects are usually described as typically quantum mechanical
in nature, but they have a perfectly natural classical analog, that I will
call the creation and annihilation functions, and that are defined as follows.

For all $\xi\in\K^\C$,
$$
a_{\mathrm{c}}(\xi): X\in \H\mapsto \bar \xi \cdot z_\Omega(X)\in\C,
$$
and
$$
a^\dagger_{\mathrm{c}}(\xi): X\in \H\mapsto  \xi \cdot
z_\Omega^\dagger(X)\in\C.
$$
Note that $a_{\mathrm{c}}(\xi)$ is anti-linear in $\xi$, whereas
$a^\dagger_{\mathrm{c}}(\xi)$ is linear. The index ``c'' stands for
``classical'', so that the notation distinguishes between the classical
creation/annihilation functions and the quantum creation/annihilation
operators, to be introduced later. A direct computation now yields
$$
\{a_{\mathrm{c}}(\xi_1), a^\dagger_{\mathrm{c}}(\xi_2)\} = -\I
\bar\xi_1\cdot \xi_2
$$
and
$$
a_{\mathrm{c}}(\xi)\circ \Phi_t = a_{\mathrm{c}}(\e^{\I\Omega
t}\xi), \quad a^\dagger_{\mathrm{c}}(\xi)\circ \Phi_t =
a^\dagger_{\mathrm{c}}(\e^{\I\Omega t}\xi).
$$
Also, for all $\eta\in\K_{-1/2}^\C$
\begin{equation}\label{eq:field}
\eta\cdot q =\frac{1}{\sqrt2}
(a_{\mathrm{c}}(\Omega^{-1/2}\bar\eta) +
a^\dagger_{\mathrm{c}}(\Omega^{-1/2} \eta))=\frac{-\I}{\sqrt2}
(a^\dagger_{\mathrm{c}}(\I\Omega^{-1/2}\eta)-a_{\mathrm{c}}(\I\Omega^{-1/2}\bar\eta)),
\end{equation}
and, similarly, for all $\eta\in\K_{1/2}^\C$
\begin{equation}\label{eq:conjugatefield}
\eta\cdot p =\frac{\I}{\sqrt2}
(a^\dagger_{\mathrm{c}}(\Omega^{1/2}\eta)-a_{\mathrm{c}}(\Omega^{1/2}\bar\eta)
).
\end{equation}
In the language of the physics literature, these two equations
express the oscilator field $\eta\cdot q$ and its conjugate field
$\eta\cdot p$ \emph{viewed as functions on phase space} in terms
of the creation and annihilation functions.

It is finally instructive to write $H$ explicitly in terms of the annihilation
and creation functions. This is easily done when $\Omega$ has pure point
spectrum, {\em i.e.} when there exists a basis of normalized eigenvectors for
$\Omega$ on $\K^\C$:
$$
\Omega \eta_i = \omega_i \eta_i,\ i\in\N.
$$
Then, from (\ref{eq:hamzomega})
\begin{equation}\label{eq:hamaadagger}
H = \frac12\sum_i \omega_i
\left(a^\dagger_{\mathrm{c}}(\eta_i)a_{\mathrm{c}}(\eta_i)
+a_{\mathrm c}(\bar\eta_i)a_{\mathrm
c}^\dagger(\bar\eta_i)\right).
\end{equation}
Note that both sides of this equation are functions on (a suitable subset of)
$\H$. Correspondingly, in quantum mechanics, both sides will be operators on
the quantum Hilbert space of states.

%%%%%%%%%%%%%%%%%%%%%%%%%%%%%%%%%%%%%%%%%%%%%%%%%%
%%%%%%%%%%%%%%%%%%%%%%%%%%%%%%%%%%%%%%%%%%%%%%%%%%
%%%%%%%%%%%%%%%%%%%%%%%%%%%%%%%%%%%%%%%%%%%%%%%%%%
%%%%%%%%%%%%%%%%%%%%%%%%%%%%%%%%%%%%%%%%%%%%%%%%%%
%%%%%%%%%%%%%%%%%%%%%%%%%%%%%%%%%%%%%%%%%%%%%%%%%%%%%%%%%%%%%%%%%%%%%%
%%%%%%%%%%%%%%%%%%%%%%%%%%%%%%%%%%%%%%%%%%%%%%%%%%%%%%%%%%%%%%%%%%%%%%
%%%%%%%%%%%%%%%%%%%%%%%%%%%%%%%%%%%%%%%%%%%%%%%%%%%%%%%%%%%%%%%%%%%%%%%
%%%%%%%%%%%%%%%%%%%%%%%%%%%%%%%%%%%%%%%%%%%%%%%%%%%%%%%%%%%%%%%%%%%%%%

\section[Quantum free harmonic systems]{The quantum theory of free harmonic systems}\label{ch:quantumfree}
\subsection[The Schr\"odinger representation]
{Finite dimensional harmonic systems: the Schr\"odinger
representation} \label{s:schrodinger} How to give a quantum
mechanical description of the classical free oscillator fields
studied in Section \ref{ch:classicalfree}? I shall proceed in two
steps. I will first recall the quantum description of a system of
a finite number of coupled oscillators, and then rewrite it in a
manner suitable for immediate adaptation to infinite dimension.

The quantum Hamiltonian for a system with $n$ degrees of freedom having a
classical Hamiltonian given by
$$
H=\frac{1}{2} p^2+ V(q),
$$
where the potential $V$ is a (smooth) real-valued function on $\R^n$ is, in
the so-called position (or Schr\"odinger) representation given by
$$
H=\frac{1}{2} P^2+ V(Q),
$$
where $P= -\I \partial /\partial x$ and $Q=x$ are the usual
momentum and position operators which are self-adjoint on there
natural domains in the ``quantum state space'' $L^2(\R^n, \d x)$.
Note that, just as in the classical description, the state space
is independent of the dynamics, which makes it easy to compare the
dynamics generated by two different Hamiltonians $H$ and $H'$,
with potentials $V$ and $V'$. To put it differently,  just as a
given point $X$ in the classical phase space $\R^{2n}$ corresponds
to the same state of the system, whatever its dynamics, so a given
$\psi$ in $L^2(\R^n)$ yields the same position and momentum
distributions for the system, whatever the dynamics to which it is
subjected.

Consequently, for an $n$-dimensional system of coupled oscillators with
classical configuration space $\K=\R^n$ and phase space $\H=\R^{2n}$ the
quantum Hamiltonian reads
$$
H=\frac{1}{2} (P^2+ Q\cdot \Omega^2 Q).
$$
Unfortunately, these expressions stop making sense when $\K$ is an
infinite dimensional space, in particular since it is not possible
to make sense out of $L^2(\K)$ in that case. So to describe the
quantum mechanics of infinite dimensional harmonic systems, I will
first rewrite the above Hamiltonian differently, in a manner
allowing for immediate generalization to infinite dimension. This
rewriting is, as we shall see, very analogous to the rewriting of
the classical mechanics on $\Kc=\C^n$, explained in Sect.
\ref{s:complexstructure}, and is therefore also affected by the
interpretational difficulties mentioned in Sect. \ref{s:physint1}.
It is nevertheless very efficient and essential.

Let's define, for any $\xi\in \Kc=\C^n$, the so-called creation and
annihilation operators
\begin{equation}\label{eq:concreteaadag}
\tilde a(\xi) =\overline\xi\cdot\frac{1}{\sqrt 2}(\Omega^{1/2}Q
+\I \Omega^{-1/2}P),\quad \tilde a^\dagger(\xi)
=\xi\cdot\frac{1}{\sqrt 2}(\Omega^{1/2}Q  -\I  \Omega^{-1/2}P).
\end{equation}
Note that those are first order differential operators, and that they depend
on $\Omega$, although the notation does not bring this dependence out. One
checks easily that
\begin{equation}\label{eq:ccr}
\left[ \tilde a(\xi), \tilde a^\dagger(\xi') \right] = \overline \xi \cdot
\xi',
\end{equation}
all other commutators vanishing. In addition, for any
$\eta\in\C^n$,
\begin{equation}\label{eq:fieldq}
\eta\cdot Q =\frac{1}{\sqrt2} (\tilde a(\Omega^{-1/2}\bar\eta) +
\tilde a^\dagger(\Omega^{-1/2} \eta)),
\end{equation}
and, similarly,
\begin{equation}\label{eq:conjugatefieldq}
\eta\cdot P =\frac{\I }{\sqrt2} (\tilde
a^\dagger(\Omega^{1/2}\eta)-\tilde a(\Omega^{1/2}\bar\eta) ).
\end{equation}
The analogy of this and of the rest of this section with the developments of
Sect. \ref{s:creationclassical} should be self-evident. In particular, it is
clear that the creation and annihilation operators are the ``quantization'' of
the creation and annihilation functions $a_{\mathrm{c}},
a^\dagger_{\mathrm{c}}$ introduced earlier.

Furthermore, let $\eta_i\in\Kc=\C^n, i=1\dots n$ be an orthonormal
basis of eigenvectors of $\Omega^2$ with eigenvalues
$\omega_1^2\leq \omega_2^2\leq \dots \leq \omega_n^2$. Then it is
easily checked that
$$
H=\frac12\sum_{i=1}^n \omega_i\left(\tilde a^\dagger(\eta_i)
\tilde a(\eta_i) + \tilde a(\bar\eta_i)\tilde
a^\dagger(\bar\eta_i)\right) = \sum_{i=1}^n \omega_i \tilde
a^\dagger(\eta_i) \tilde a(\eta_i) + \frac12 \sum_{i=1}^n
\omega_i.
$$
The spectral analysis of $H$ is now straightforwardly worked out, and
described in any textbook on quantum mechanics. Let me recall the essentials.

It is first of all readily checked that there exists a unit vector
$|0,\Omega\rangle$ in $L^2(\R^n)$ (unique up to a global phase), for which
$$
\tilde a(\xi)|0, \Omega\rangle =0, \forall \xi\in\C^n.
$$
This common eigenvector of all the annihilation operators $\tilde a(\xi)$ is
called the ``vacuum''. Remark that, as a vector in $L^2(\R^n)$, the vacuum
$|0, \Omega\rangle$ obviously depends on $\Omega$. One has indeed very
explicitly
\begin{equation}\label{eq:groundstate}
\langle x|0,\Omega\rangle = \frac{(\mathrm{det}\Omega)^{1/4}}{\pi^{n/4}}
\exp-\frac{1}{2}x\cdot\Omega x.
\end{equation}

Clearly $H|0, \Omega\rangle=\frac12 \sum_{i=1}^n \omega_i
|0,\Omega\rangle$, so that the vacuum $|0,\Omega\rangle$ is
actually the ground state of $H$.  Writing for brevity $\tilde
a_i=\tilde a(\eta_i), \tilde a^\dagger_i = \tilde
a^\dagger(\eta_i)$ it follows (after some work) that the vectors
\begin{equation}\label{eq:oscillatorbasis}
\frac{1}{\sqrt{m_1!m_2!m_3!\dots m_n!}}\left(\tilde
a^\dagger_{1}\right)^{m_1}\left(\tilde a^\dagger_{2}\right)^{m_2}\left(\tilde
a^\dagger_{3}\right)^{m_3}\dots \left(\tilde
a^\dagger_{n}\right)^{m_n}|0,\Omega\rangle,
\end{equation}
for all possible choices $(m_1,\dots, m_n)\in\N^n$ form an orthonormal basis
of eigenvectors for $H$.

Note that the position and momentum distributions of the ground state
evidently depend on $\Omega$ and are in fact not totally trivial to compute,
despite the apparent simplicity of the Gaussian expression above. Indeed, if
you want to know, for example,
 $\langle 0,\Omega|Q_7^2|0,\Omega\rangle$ you actually need to be able to diagonalize $\Omega^2$ explicitly,
  and you need in particular an explicit description of the normal modes.
  This can be done in simple cases, such as the oscillator ring, but not in general.

One can also introduce the ``number operator''
$$
\tilde N=\sum_{i=1}^n \tilde a^\dagger_i \tilde a_i,
$$
which commutes with $H$. The spectrum of $\tilde N$ is easily seen to equal to
$\N$. Writing ${\mathcal{E}}_m$ for the eigenspace of $\tilde N$ with
eigenvalue $m$, one has evidently
\begin{equation}\label{eq:l2sum}
L^2(\R^n)= \sum_{m\in\N}^\oplus \mathcal{E}_m.
\end{equation}
Each vector in (\ref{eq:oscillatorbasis}) is readily checked to be  an
eigenvector of $\tilde N$ with eigenvalue $\sum_{k=1}^n m_k$. The preceding
considerations will be the starting point for an equivalent reformulation of
the quantum theory of finite dimensional oscillator systems in a manner
suitable for generalization to infinite dimensional systems. This
reformulation is based in an essential manner on the notion of  Fock space,
which I therefore first briefly recall in the next section.

%%%%%%%%%%%%%%%%%%%%%%%%%%%%%%%%%%%%%%%%%%
%%%%%%%%%%%%%%%%%%%%%%%%%%%%%%%%%%%%%%%%%%%%
%%%%%%%%%%%%%%%%%%%%%%%%%%%%%%%%%%%%%%%%%%%5
%%%%%%%%%%%%%%%%%%%%%%%%%%%%%%%%%%%%%%%%%%%5

\subsection{Fock spaces}\label{s:fock}
The basic theory of symmetric and anti-symmetric Fock spaces can
be found in many places (\cite{rs} \cite{br1} are two examples)
and I will not detail it here, giving only the bare essentials,
mostly for notational purposes. More information on this subject
can also be found in the contribution of Jan Derezinski in this
volume \cite{d}.

Let $\V$ be a complex Hilbert space, then the
 Fock space $\eff(\V)$ over $\V$ is
$$
\eff(\V) =\overline{\oplus_{m\in\N} \eff_{m}(\V)},
$$
where $\eff_m(\V)$ is the $m$-fold  tensor product of $\V$ with itself.
Moreover $\eff_0(\V)=\C$. An element $\psi\in\eff(\V)$ can be thought of as a
sequence
$$\psi=(\psi_0, \psi_1,\dots,\psi_m,\dots),$$
where $\psi_m\in\eff_m(\V)$. I will also use the notation
$\eff^{\mathrm{fin}}(\V)=\oplus_{m\in\N} \eff_{m}(\V)$, which is
the dense subspace of $\eff(\V)$ made up of elements of the type
$$\psi=(\psi_0, \psi_1,\dots,\psi_N,0,0,\dots)$$
for some integer $N\geq 0$. Don't confuse $\eff_0(\V)$ with
$\eff^{\mathrm{fin}}(\V)$! Elements of $\eff^{\mathrm{fin}}(\V)$
will be referred to as states \emph{with a finite number of
quanta}, a terminology that I will explain later.

I will freely use the Dirac notation for Hilbert space calculations. So I will
write $|\psi\rangle\in\eff(\V)$ as well as $\psi\in\eff(\V)$, depending on
which one seems more convenient at any given time. Also, when no confusion can
arise, I will write $\eff_m=\eff_m(\V)$.

Let $\P_m$ be the permutation group of $m$ elements, then for each
$\sigma\in\P_m$, we define the unitary operator $\hat\sigma$ on $\eff_m(\V)$
by
$$
\hat \sigma \xi_1\otimes \xi_2\otimes\dots\otimes \xi_m =
\xi_{\sigma^{-1}(1)}\otimes \xi_{\sigma^{-1}(2)}\otimes\dots
\otimes \xi_{\sigma^{-1}(m)},
$$
($\xi_j\in\C^n, j=1,\dots, m$) and the projectors
$$
P_{+,m} =\frac{1}{m!}\sum_{\sigma\in\P_m}\hat \sigma,\qquad P_{-,m}
=\frac{1}{m!}\sum_{\sigma\in\P_m}\mathrm{sgn}(\sigma)\hat \sigma.
$$
Now we can define the (anti-)symmetric tensor product as
$$
\eff_{m}^{\pm}(\V) = P_{\pm, m}\eff_{m}(\V)
$$
whereas the (anti-)symmetric Fock space $\eff^{\pm}(\V)$ over $\V$ is
$$
\eff^{\pm}(\V)=\overline{\oplus_{m\in\N} \eff^{\pm}_{m}(\V)}.
$$
Introducing the projector $P_{\pm}=\sum_{m\in\N} P_{\pm,m}$, we also have
$$
\eff^{\pm}(\V)=P_{\pm}\eff(\V)\quad\mathrm{and}\quad
\eff^{\mathrm{fin},\pm}(\V)=P_{\pm}\eff^{\mathrm fin}(\V).
$$
One refers to $\eff^+(\V)$ as the symmetric or bosonic Fock space and to
$\eff^-(\V)$ as the anti-symmetric or fermionic Fock space. I will only deal
with the former here.

Computations in Fock space are greatly simplified through the use of
``creation'' and ``annihilation'' operators, which are abstract versions of
the operators $\tilde a(\xi)$ and $\tilde a^\dagger(\xi)$ introduced in
Sect. \ref{s:schrodinger}.

Define, for any $\xi\in\V$,
$$
d(\xi) \xi_1\otimes \xi_2\dots\otimes \xi_m= (\overline \xi\cdot \xi_1)\
\xi_2\otimes\dots\otimes \xi_m.
$$
This extends by linearity and yields a well-defined bounded operator from
$\eff_m$ to $\eff_{m-1}$ which extends to a bounded operator on all of
$\eff(\V)$, denoted by the same symbol.

Note that I use the notation $\overline \xi\cdot \eta$  for the inner product
on the abstract space $\V$ because in the applications in these notes $\V$
will be $\Kc$, in which case this notation is particularly transparent. Of
course, on a general abstract $\V$, there is no natural definition of ``the
complex conjugate $\overline \xi$'', but that does not mean we can't use
$\overline \xi\cdot \eta$ as a notation for the inner product.

One has $\parallel d(\xi)\parallel = \parallel \xi \parallel$. Similarly,
define
$$
c(\xi)\xi_1\otimes \xi_2\dots\otimes \xi_m= \xi\otimes\xi_1\otimes
\xi_2\otimes\dots\otimes \xi_m.
$$
This again yields a well-defined bounded operator from $\eff_m$ to
$\eff_{m+1}$ which extends to a bounded operator on all of $\eff(\V)$, denoted
by the same symbol. One has $\parallel c(\xi)\parallel = \parallel \xi
\parallel$ and $d(\xi)^* = c(\xi)$.

Introducing the self-adjoint ``number operator'' $N$ by
$$
N\psi = (0, \psi_1, 2\psi_2, \dots m\psi_m\dots ),
$$
we can then define, on $\eff^{\mathrm{fin}}$,
$$
a_\pm(\xi) = P_{\pm} \sqrt{N+1} d(\xi)P_{\pm},\qquad\mathrm{and}\qquad
a_\pm^\dagger(\xi) = P_{\pm} \sqrt{N} c(\xi)P_{\pm}.
$$
The $a_\pm(\xi)$ are called ``annihilation operators'' and the
$a^\dagger_\pm(\xi)$ creation operators. I will think of
$a_-(\xi)$ as an operator on $\eff^-$ and of $a_+(\xi)$ as an
operator on $\eff^+$. Direct computation (on $\eff^{\mathrm fin}$,
for example) yields the following crucial commutation and
anti-commutation relations between those operators:
\begin{equation}\label{eq:ccrabstract}
[a_+(\xi_1), a_+(\xi_2)]=0=[a_+^\dagger(\xi_1), a_+^\dagger(\xi_2)],\qquad
 [a_+(\xi_1), a_+^\dagger(\xi_2)] = \overline{\xi_1}\cdot\xi_2,
\end{equation}
Those  are referred to as the canonical
commutation relations or CCR. You should compare (\ref{eq:ccrabstract})
to (\ref{eq:ccr}) and be amazed.

Working in the bosonic Fock space $\eff^+$ and using the above relations one
establishes through direct computation that
$$
\sqrt{m!}P_+\xi_1\otimes\xi_2\otimes\dots\otimes\xi_m =
a_+^\dagger(\xi_1)a_+^\dagger(\xi_2)\dots a_+^\dagger(\xi_m)|0\rangle.
$$
Here I introduced the notation $|0\rangle =
(1,0,0,\dots)\in\eff_0\subset\eff$. This vector is usually referred to as the
Fock vacuum or simply as the vacuum. It can be characterized as being the
unique vector in $\eff^+$ for which
$$
a_+(\xi)|\psi\rangle = 0, \qquad \forall \xi\in\V.
$$
For explicit computations and in order to understand the physics literature,
it is a Good Thing to have a convenient basis at hand. So suppose you have an
orthonormal basis $\eta_j$ of $\V$ (with $j=1, 2, \dots \mathrm{dim} \V$).
Then you can define, for any positive integer $k\leq \mathrm{dim}\V$ and for
any choice of $(m_1, m_2, m_3, \dots, m_k)\in\N^k$, the vector
\begin{equation}\label{eq:fockbasis}
{|m_1, m_2, m_3,\dots, m_k\rangle :=(m_1!m_2!\dots
m_k!)^{-1/2}}\times \qquad\qquad\qquad\qquad\qquad\qquad
\end{equation}
$$
\left(a^\dagger_+(\eta_1)\right)^{m_1}
\left(a_+^\dagger(\eta_2)\right)^{m_2}\left(a_+^\dagger(\eta_3)\right)^{m_3}\dots\left(a_+^\dagger(\eta_k)\right)^{m_k}|0\rangle.
$$
Those vectors are now easily checked to form an orthonormal basis of $\eff^+$.
The numbers $m_j$ are often referred to as the ``occupation numbers'' of the
states $\eta_j$. Note that each of them is an eigenvector of the number
operator with eigenvalue given by $\sum_{j=1}^k m_j$.

It is a good exercise to prove that $N_+=P_+NP_+$, the restriction of the
number operator to $\eff^+$ can be written
$$
N_+=\sum_j a_+^\dagger(\eta_j)a_+(\eta_j).
$$

If $U$ is a unitary operator on $\V$, the unitary operator
$\Gamma(U)$ on $\eff^+$ is defined as $\otimes_{k=1}^m U$ when
restricted to $\eff^+_m$. When $A$ is a self-adjoint operator on
$\V$, $\d\Gamma(A)$ is the self-adjoint operator on $\eff^+$
defined as
$$
A\otimes \bbbone \otimes\dots\otimes \bbbone + \bbbone\otimes A
\otimes\dots\otimes \bbbone + \dots + \bbbone\otimes  \dots\otimes
\bbbone\otimes A
$$
on (a suitable domain) in $\eff_m^+$, for each $m>0$. Also
$\d\Gamma(A)\eff_0^+=0$. It is a good exercise to check that, if
$A$ has a basis of eigenvectors
$$
A\eta_j=\alpha_j\eta_j
$$
then
$$
\d\Gamma(A)=\sum_i \alpha_i\ a_+^\dagger(\eta_i)a_+(\eta_i).
$$

%%%%%%%%%%%%%%%%%%%%%%%%%%%%%%%%%%%%%%%%%%%%%%%%%%
%%%%%%%%%%%%%%%%%%%%%%%%%%%%%%%%%%%%%%%%%%%%%%%%

\subsection[Fock  representation: finite dimensional fields]
{The Fock  representation: finite dimensional fields}\label{s:focrepfin} It is
now straightforward to reformulate the quantum description of the oscillator
system in Sect. \ref{s:schrodinger} as follows. First of all, in view of
(\ref{eq:l2sum}) and the considerations of the previous section,
 it is clear that there
exists a unitary map $T_\Omega$
$$
T_\Omega : L^2(\R^n)\to\eff^+(\C^n)
$$
satisfying
$$
T_\Omega\mathcal{E}_m=\eff^+_m(\C^n), \quad T_\Omega H
T_\Omega^{-1} = \d\Gamma(\Omega)
 +\frac12\sum_{i=1}^n \omega_i,
$$
and
$$
T_\Omega \tilde a(\xi) T_\Omega^{-1}=a_+(\xi),
$$
for all $\xi\in\C^n$. In fact, quite explicitly, one has, for all
$\xi_1,\xi_2,\dots\xi_m\in\C^n$,

\begin{eqnarray*} T_\Omega: \tilde a^\dagger(\xi_1)\dots \tilde
a^\dagger(\xi_m)|0, \Omega\rangle&\in&\mathcal{E}_m\subset
L^2(\R^n)\\
&\mapsto& a_+^\dagger(\xi_1)\dots a_+^\dagger(\xi_m)|0\rangle\in
\eff_m^+(\C^n)\subset\eff^+(\C^n).
\end{eqnarray*}
The unitary map $T_\Omega$ transports each object of the theory from
$L^2(\R^n)$ to the symmetric Fock space over $\C^n$  and provides in this
manner an equivalent quantum mechanical description of the oscillator system,
that goes under the name of Fock representation.

Note that in the left hand side of the above equations, the operators $\tilde
a(\xi)$ or $\tilde a^\dagger(\xi)$ are the concrete differential operators on
$L^2(\R^n)$ that were defined in (\ref{eq:concreteaadag}) and that depend
explicitly on $\Omega$. In the right hand side, you find the abstract creation
and annihilation operators defined in Sect. \ref{s:fock}. Note that those do
not depend on $\Omega$ at all. Similarly, the ground state vector $|0,
\Omega\rangle$ of $H$ appearing in the left hand side is of course
$\Omega$-dependent, whereas the Fock vacuum $|0\rangle$ in the right hand side
is not. This is somewhat paradoxical. Indeed, since the vacuum is the ground
state of the Hamiltonian, should it not depend on this Hamiltonian? The answer
to this conundrum goes as follows, and is very similar to the discussion in
Sect. \ref{s:physint1} in the classical context. Recall that it is customary
to say that each physical state of the system is represented by a vector in a
Hilbert space. Consider for example the vacuum vector $|0\rangle$ in Fock
space. To find out to which physical state of the system it corresponds, one
has to compute the expectation value of physical observables in this state.
Now, for a system of coupled oscillators, the most relevant observables are
arguably the coordinates of position and momentum. In view of
(\ref{eq:fieldq}) and (\ref{eq:conjugatefieldq}) it is now clear that
\begin{equation}\label{eq:fieldq2}
T_\Omega\eta\cdot QT_\Omega^{-1} =\frac{1}{\sqrt2}
(a_+(\Omega^{-1/2}\overline\eta) + a_+^\dagger(\Omega^{-1/2}
\eta)),
\end{equation}
and, similarly,
\begin{equation}\label{eq:conjugatefieldq2}
T_\Omega\eta\cdot PT_\Omega^{-1} =\frac{i}{\sqrt2}
(a_+^\dagger(\Omega^{1/2}\eta)-a_+(\Omega^{1/2}\overline\eta) ).
\end{equation}
I will in the following  not hesitate to write $T_\Omega\eta\cdot
QT_\Omega^{-1}=\eta\cdot Q$ and $T_\Omega\eta\cdot PT_\Omega^{-1}=\eta\cdot
P$, in agreement with the usual convention that consists of not making the
identification operator $T_\Omega$ notationally explicit. But it is now clear
that, contrary to what happens on $L^2(\R^n)$, the explicit expression of the
position and momentum observables as operators on Fock space depends  on the
dynamics, via $\Omega$!  Hence the expectation values of those operators, and
of polynomial expressions in these operators will also depend on $\Omega$. In
this sense, {\em the same mathematical object}, namely the vector
$|0\rangle\in\eff^+(\C^n)$ {\em corresponds to a different physical state of
the system} of $n$ coupled oscillators for different choices of $\Omega$, {\em
i.e.} of the spring constants. Also, the \emph{same physical quantity}, such
as the displacement of the seventh oscillator, is represented by a different
mathematical operator, namely the operator in the right hand side of
(\ref{eq:fieldq2}), with $\eta(j)=\delta_{j7}$. In particular, if you are
interested in the mean square displacement of the seventh oscillator when the
system is in the ground state, \emph{i.e.} $\langle 0| Q_7^2|0\rangle$, you
will need a detailed spectral analysis of $\Omega$ and in particular a good
understanding of the spatial distribution of its normal modes over the $n$
degrees of freedom of the system, as I already pointed out. The result you
find will of course depend on $\Omega$.

In the same manner, any other given fixed vector in the Fock space, such as
for example a state of the form $a_+^\dagger(\xi)|0\rangle$, for some fixed
choice of $\xi\in\C^n$, represents a {\em different} physical state depending
on $\Omega$.

In short, the interpretation of a given vector in Fock space as a state of a
physical system depends on the dynamics of the system under consideration
because the representation of the  physical observables of the system by
operators on Fock space is dynamics dependent.

To avoid confusion, these simple remarks need to be remembered when dealing
with the infinite dimensional theory, where only the Fock representation
survives. In particular, the name ``vacuum vector'' or ``vacuum state'' given
to the Fock vacuum conveys the wrong idea that, somehow, when the system state
is represented by this vector, space is empty, there is ``nothing there'' and
therefore this state should have trivial physical properties that in fact
should be independent of the system under consideration and in particular of
the dynamics.

%%%%%%%%%%%%%%%%%%%%%%%%%%%%%%%%%%%%%%%
%%%%%%%%%%%%%%%%%%%%%%%%%%%%%%%%%%%%%%%
%%%%%%%%%%%%%%%%%%%%%%%%%%%%%%%%%%%%%%
%%%%%%%%%%%%%%%%%%%%%%%%%%%%%%%%%%%%%%%%

\subsection[Fock representation: general free fields]
{The Fock representation: general free
fields}\label{s:fockreprgeneral} Summing up, we have  now
reformulated the quantum mechanical description of a finite
dimensional coupled oscillator system in a way that will be seen
to carry over immediately -- with only one moderate change -- to
the infinite dimensional case. Indeed, given a free oscillator
field determined by $\K$ and $\Omega$, it is now perfectly natural
to choose as the quantum Hilbert space of such a system the Fock
space $\eff^+(\K^\C)$, and as quantum Hamiltonian
$H=\d\Gamma(\Omega)$. Note that this is a positive operator and
that the Fock vacuum is its ground state, with eigenvalue $0$.
Proceeding in complete analogy with the finite dimensional case,
the quantization of the classical creation and annhilition
functions $a_\c(\xi), a^\dagger_\c(\xi)$ are the creation and
annihilation operators $a_+(\xi), a_+^\dagger(\xi)$. In terms of
those the quantized fields and their conjugates are then {\em
defined} precisely as before ($\eta\in\K^\C_{-1/2}$):
\begin{equation}\label{eq:fieldq3}
\eta\cdot Q := \frac{1}{\sqrt2} (a_+(\Omega^{-1/2}\eta) +
a_+^\dagger(\Omega^{-1/2} \eta)),
\end{equation}
and, similarly ($\eta\in\K^\C_{1/2}$),
\begin{equation}\label{eq:conjugatefieldq3}
\eta\cdot P := \frac{\I }{\sqrt2}
(a_+^\dagger(\Omega^{1/2}\eta)-a_+(\Omega^{1/2}\overline\eta) ).
\end{equation}
It is often convenient to think of ``the field $Q$'' as the map that
associates to each $\eta\in\K_{-1/2}$ the self-adjoint operator in the right
hand side of (\ref{eq:fieldq3}), and similarly for ``the conjugate field
$P$'', defined on $\K_{1/2}$. With this language, the field operator
$\eta\cdot Q$ is the value of the field $Q$ at $\eta\in\K_{-1/2}$. This
notation is reasonable since the field is a linear function of its argument.

The moderate change to which I referred to above is the fact that,
if I compare the above quantization prescription for the case
$\K=\R^n$ to the one of Sect. \ref{s:schrodinger} and Sect.
\ref{s:focrepfin}, then it is clear that I substracted from the
Hamiltonian the ``zero-point energy'', $\sum_{i=1}^n\omega_i$.  It
is argued in all quantum field theory texts that this constitutes
an innocuous change, for two distinct reasons. First,  adding a
constant to the Hamiltonian does not change the dynamics in any
fundamental way. Second only energy differences count in physics,
so tossing out an additive constant in the definition of the
energy should not change anything fundamentally. As a result,
since the expression $\sum_{i=1}^n\omega_i$ makes no sense in
general in infinite dimensions, where it is formally typically
equal to $+\infty$, it seems like a good idea to toss it out from
the very beginning! This means you calibrate the energy so that
the ground state of the system, which is represented by the Fock
vacuum, has zero {\em total} energy, independently of $\Omega$,
and leads to the choice of $H=\d\Gamma(\Omega)$ as the
Hamiltonian.

While this is the reasoning found in all physics and mathematical physics texts the tossing out of the zero-point
energy is not such an innocent operation after all. For the physics of the zero-point
energy, I refer to  \cite{mi}. See also \cite{db2} for further comments.

It is instructive to compute the evolution of the field and the conjugate
field under the dynamics. Since
$$
\e^{ -\I  H t} = \Gamma(\e^{ -\I \Omega t}),
$$
it is easy to check that
$$
\e^{\I  H t} a_+(\xi)\e^{ -\I  H t} = a_+(\e^{\I  \Omega t}\xi).
$$
Define then the evolved field $Q(t)$ as the map that associates to each
$\eta\in\K_{-1/2}$ the self-adjoint operator $\eta\cdot Q(t)$ defined as
follows:
$$
\eta\cdot Q(t)\equiv  \e^{\I H t}(\eta\cdot Q)\e^{ -\I  H t}.
$$
A simple computation then yields
$$
\eta\cdot Q(t) = \frac{1}{\sqrt2} (a_+(\Omega^{-1/2}\e^{\I \Omega
t}\overline \eta) + a_+^\dagger(\Omega^{-1/2} \e^{\I \Omega
t}\eta)).
$$
Hence
$$
\frac{\d^2}{\d t^2}\eta\cdot Q(t) = -\Omega^2\eta\cdot Q(t).
$$
One defines similarly $\eta\cdot P(t)$, which obeys the same
equation. In fact, $\eta\cdot Q(t)$ and $\eta\cdot P(t)$ are
operator-valued solutions of this with $\eta\cdot Q(t)$ satisfying
the \emph{equal time commutation relations}. They are called the
Heisenberg field and conjugate field in the physics literature.

We are now in a position to further study these systems, a task I turn to
next. First, a word on the ``particle interpretation of the field states'' is
in order.

%%%%%%%%%%%%%%%%%%%%%%%%%%%%%%%%%%%%%%%%%%%%%%%%
%%%%%%%%%%%%%%%%%%%%%%%%%%%%%%%%%%%%%%%%%%%%%%%%%%
%%%%%%%%%%%%%%%%%%%%%%%%%%%%%%%%%%%%%%
%%%%%%%%%%%%%%%%%%%%%%%%%%%%%%%%%%
%%%%%%%%%%%%%%%%%%%%%%%%%%%%%%%%%%%%%%%%
\subsection{Particle interpretation of the field states}\label{s:parint}
Physicists refer to $\eff_{m}$ as the $m$ particle
 sector of the Fock space ($m\geq1$) and to $\eff_0$ as the vacuum sector.
 This terminology comes from the following remark. As any beginners' text in quantum mechanics will
tell you, whenever the quantum Hilbert space of a single particle (or a single
system) is $\V$, the Hilbert space of states for $m$ (identical) particles (or
systems) is the $m$-fold tensor product of $\V$. The simplest case is the one
where $\V=L^2(\R^d)$. Then the $m$-fold tensor product can be naturally
identified with $L^2(\R^d\times\dots\times\R^d=\R^{dm})$, which is isomorphic
to $\otimes_m \V$. The same quantum mechanics course will teach you that, when
the particles are indistinguishable, the state space needs to be restricted
either to the symmetric or anti-symmetric tensor product. In the first case,
which is the one we are dealing with here, the particles are said to be
bosons, otherwise they are fermions.  In the case where $\V=L^2(\R^d)$, the
$m$-fold symmetric tensor product of $\V$ consists of all symmetric
$L^2$-functions of $m$ variables.

The above considerations  suggest that, {\em conversely}, whenever the quantum
state space of a physical system turns out to be a Fock space over some
Hilbert space $\V$, one may think of $\V$ as a one-particle space, and of
$\eff_{m}(\V)$ as the corresponding $m$-particle space. An arbitrary state of
the system can then be thought of as a superposition of states with $0$, $1$,
$2$, \dots $m$, \dots particles. These ideas emerged very quickly after the
birth of quantum mechanics, as soon as physicists attacked the problem of
analyzing the quantum mechanical behaviour of systems with an infinite number
of degrees of freedom, such as the electromagnetic field. The Fock space
structure of the Hilbert space of states describing the {\em field}
immediately lead to such an interpretation in terms of \emph{particles}. For
the electromagnetic field, the particles were baptized ``photons'', and in
complete analogy, the quantum mechanical description of lattice vibrations in
solid state physics lead to the notion of ``phonons''. The idea that one can
associate a particle interpretation to the states of a Fock space is further
corroborated by the observation that those states carry energy and momentum in
``lumps''. This can be seen as follows. Suppose, in our notations, that
$\Omega$ has a pure point spectrum:
$$
\Omega\eta_j=\omega_j\eta_j, \ j\in\N.
$$
Then the quantum Hamiltonian is
$$
H=\d\Gamma(\Omega)=\sum_j \omega_j a^\dagger_j a_j,
$$
where I wrote $a^\dagger_j=a^\dagger(\eta_j)$. Note that I have
dropped the index $+$ on the creation and annihilation operators,
a practice that I shall stick to in what follows since I will at
any rate be working on the symmetric Fock space all the time.
 Now consider for example the state
$$
a^\dagger_1(a^\dagger_5)^3a^\dagger_{10}|0\rangle.
$$
This is a $5$-particle state, and an eigenvector of the Hamiltonian with
eigenvalue $\omega_1 + 3\omega_5 +\omega_{10}$. It is natural to think of it
intuitively as being a state ``containing'' $3$ particles of energy
$\omega_5$, and one particle of energy $\omega_1$ and $\omega_{10}$ each.
Similarly, in translationally invariant systems, such states can be seen to
carry a total momentum which is the sum of ``lumps'' of momentum corresponding
to its individual constituents. Of course, the particle interpretation of the
states of the field is a very important feature of the theory since it is
essential for the interpretation of high energy experiments, and so it has
quite naturally received a lot of attention.

Despite its undeniable value, the suggestive interpretation of the states of
Fock space in terms of particles may   lead (and has lead) to some amount of
confusion and has to be taken with a (large) grain of salt. Some of those
problems seem to have been brought out clearly only when physicists started to
investigate  quantum field theory on curved space-times. A critical discussion
of this issue can be found throughout \cite{fu}. Although Fulling does adopt
the second quantization viewpoint, he stresses repeatedly the need to escape
``from the tyranny of the particle concept'' in order to ``come to a
completely field theoretic understanding of quantum field theory.'' Similarly,
Wald, who does indeed adopt a field theoretic viewpoint throughout  in
\cite{wa}, gives a critical analysis of the merits and limitations of the
particle concept in quantum field theory. He actually stresses the need to
``unlearn'' some of the familiar concepts of quantum field theory on flat
space times to understand the curved space time version of the theory.

There are in fact several sources of problems with the particle interpretation
of the states in quantum field theory. The first one was already hinted at in
Sect. \ref{s:focrepfin}:  the use of the word ``vacuum'' to describe the
ground state of the system invites one to think that when the system is in
this state, there is ``nothing there''. Actually, one may be tempted to think
the system itself is simply not there! But to see that makes no sense, it is
enough to think of an oscillator lattice. Certainly, when this system is in
its ground state, all oscillators are there! It is just that the system is not
excited, so there are no ``particles'' in the above (Fock space) sense of the
word, and this in spite of the fact that the mechanical particles making up
the lattice are certainly present.   Also, if one thinks of the vacuum state
as empty space, it becomes impossible to understand how its properties  can
depend on the system considered  via $\Omega$. In fact, it is quite baffling
to think ``empty space'' could have any properties at all. In particular, the
mean square displacement of the field, for example, given by
$$
\langle 0|(\eta\cdot Q)^2|0\rangle
$$
is a function of $\Omega$, as is easily seen even in finite dimensional oscillator
systems. This
quantity is an example of a so-called ``vacuum fluctuation''.  Of course, for
systems with a finite number of degrees of freedom, we find this phenomenon
perfectly natural, but if you study the Klein-Gordon field, for example, and
call the ground state the vacuum, you end up being  surprised to see vacuum
expectation values depend on the mass of particles that are not there!

A second source of confusion is that the notion of ``particle'' evokes a
localized entity, carrying not only momentum and energy, but that one should
also be able to localize in space, preferably with the help of a position
operator. I will show in Section \ref{ch:locobssta} that there is no
reasonable notion of ``position'' that can be associated to the one-particle
states of Fock space, contrary to what happens in the usual non-relativistic
quantum mechanics of systems with a finite number of particles. In particular,
there is no reasonable ``position operator''.  This has nothing to do with
relativity, but is true for large classes of $\Omega$ and in particular for
all examples given so far. So even if the particles of field theory share a
certain number of properties with the usual point particles of classical and
quantum mechanical textbooks, they have some important features that make them
quite different. They are analogous objects, but not totally similar ones.
This, I will argue, has nothing to do either with special or general
relativity, but is clear if one remembers systematically the analogy with
finite dimensional oscillator systems.

As a constant reminder of the fact that the so-called
particles of quantum field theory are nothing but excitations of its ground
state,  it  is  a good idea to use the older physics terminology and to talk
systematically of
 ``quasi-particles'', ``quanta'', ``field quanta'' or of ``elementary excitations of the field'' rather than simply of particles when describing the states of Fock space. I will adhere as much as possible
to this prudent practice.

Moreover, when testing your understanding of a notion in quantum field theory,
try to see what it gives for a finite system of oscillators. If it looks funny
there, it is likely to be a bad idea to use it in the infinite dimensional
case.

The remaining parts of this section develop material that will be needed in
Section \ref{ch:locobssta}. It is perhaps a good idea to start reading the
latter, coming back to this material only as I refer to it.

%%%%%%%%%%%%%%%%%%%%%%%%%%%%%%%%%%%%%%%%%%%%%%%%%%
%%%%%%%%%%%%%%%%%%%%%%%%%%%%%%%%%%%%%%%%%%%%%%%%%%

\subsection{Weyl operators and coherent states}\label{s:weyloperators}
Given a Hilbert space $\V$ and the corresponding symmetric Fock
space $\eff^+(\V)$, we can first define, for any $\xi\in\V$, the
{\em Weyl operator}
$$
W_{\mathrm{F}}(\xi) = \e^{a^\dagger(\xi)-a(\xi)}.
$$
 A coherent state is then defined as a
vector of $\eff^+(\V)$ of the form
$$
|\xi\rangle\stackrel{\mathrm{def}}{=} \Wf(\xi)|0\rangle,
$$
for some $\xi\in\V$. Note that the map
$$
\xi\in\V\mapsto |\xi\rangle\in \eff^+(\V)
$$
provides a {\em nonlinear} imbedding of $\V$ into $\eff^+(\V)$ which is not to
be confused with the trivial linear imbedding $\V\cong\eff_1(\V)\subset
\eff^+(\V)$. Given an arbitrary $0\not=\psi\in\V$, one can likewise consider
the family $\Wf(\xi)\psi$, and those vectors are also referred to as a family
of coherent states.

Coherent states play an important role in the semi-classical analysis of
quantum systems and in various branches of theoretical physics \cite{ksk}
\cite{pe}. We describe them here in the abstract context of symmetric Fock
spaces. They are very simple objects to define but nevertheless have an
seemingly inexhaustable set of interesting properties. I will only mention
those I need.

To compute with the coherent states, we need  a number of formulas that are
listed below and that can all be obtained easily, if one remembers first of
all that, if $A$ and $B$ are bounded operators so that $C=[A,B]$ commutes with
both $A$ and $B$, then
$$
\e^{A+B} = \e^A \e^B \e^{-\frac{C}{2}}, C=[A,B].
$$
Computing with $a^\dagger(\xi)$ and $a(\xi)$ as if they were bounded
operators, all formulas below follow from this and some perseverance in
computing. Taking care of the domain problems to make them completely rigorous
is tedious but character building and can be done using the techniques
described in \cite{br2} or \cite{rs}. First of all, we have, for all $\xi_1,
\xi_2\in\V$,
$$
\Wf(\xi_1)\Wf(\xi_2) = \Wf(\xi_1+\xi_2)\e^{ -\I {\mathrm{Im}}
(\overline\xi_1\cdot \xi_2)}.
$$
As a result
$$
\Wf(\xi_1)\Wf(\xi_2) = \Wf(\xi_2) \Wf(\xi_1)\e^{-2\I{\mathrm{Im}}
(\overline\xi_1\cdot \xi_2)},
$$
and
$$
\left[\Wf(\xi), \Wf(\xi')\right] = \Wf(\xi')\Wf(\xi)\left(\e^{\I
2\mathrm{Im}(\overline\xi'\cdot\xi)}-1\right).
$$
 Furthermore
$$
\Wf(s\xi)\Wf(\xi')\Wf(t\xi)=\Wf((s+t)\xi)\Wf(\xi') \e^{2\I
t\mathrm{Im}(\overline \xi\cdot \xi')}
$$
and hence
$$
\Wf(-\zeta)\Wf(\xi)\Wf(\zeta) = \Wf(\xi) \e^{\I
2\mathrm{Im}(\overline \zeta\cdot\xi)},
$$
or
$$
\Wf(-\zeta)\Wf(\xi)\Wf(\zeta) = \e^{\left[(a^\dagger(\xi) +
\xi\cdot\overline\zeta) - (a(\xi) + \overline \xi \cdot \zeta)\right]}.
$$
One then finds
$$
\Wf(-\zeta)\left[a^{\dagger}(\xi)\right]^n\Wf(\zeta)=(a^\dagger(\xi)
+ \xi\cdot\overline\zeta)^n,\
\Wf(-\zeta)a^{n}(\xi)\Wf(\zeta)=(a(\xi) +
\overline\xi\cdot\zeta)^n.
$$
It is often convenient to write
$$
\Wf(\xi) = \e^{a^\dagger(\xi)}\e^{-a(\xi)}\e^{-\frac12\parallel
\xi\parallel^2} =\e^{-a(\xi)}\e^{a^\dagger(\xi)}\e^{\frac12\parallel
\xi\parallel^2}.
$$
Also, remark that, for all $\xi\not=0$,
$$
\parallel \Wf(\xi)- \bbbone\parallel =2\ \mathrm{and}\
\mathrm{s}-\lim_{t\to0} \Wf(t\xi)= \bbbone.
$$

Using what precedes, one easily finds the following formulas
involving the vacuum.
\begin{equation}\label{eq:vac1}
|\xi\rangle = \e^{-\frac12 \parallel
\xi\parallel^2}\e^{a^\dagger(\xi)}|0\rangle,
\end{equation}
\begin{equation}\label{eq:vac2}
\langle 0|a^n(\xi)|\xi'\rangle=  \e^{-\frac12 \parallel \xi'
\parallel^2} (\overline \xi\cdot\xi')^n,
\end{equation}
and
\begin{equation}\label{eq:vac3}
\langle \zeta| \Wf(\xi)|\zeta\rangle = \e^{-\frac12\parallel
\xi\parallel^2}\e^{\I 2\mathrm{Im}(\overline \zeta\cdot\xi)}.
\end{equation}

%%%%%%%%%%%%%%%%%%%%%%%%%%%%%%%%%%%%%%%%%%%%%%%%%%%%
%%%%%%%%%%%%%%%%%%%%%%%%%%%%%%%%%%%%%%%%%%%%%%%%%%%%
\subsection{Observables and observable algebras}\label{s:obsobsalg}
Physically measurable quantities of a system are,  in its  classical
description, represented by
 functions on phase space.  Consider first finite dimensional systems.
 An example, in the case of an oscillator ring,
  is ``the displacement of the ninth oscillator'', represented by
 $q_9:X=(q,p)\in\H\to q_9\in\R$. Some interesting observables are represented by linear functions (such
as position and momentum) or by quadratic functions (such as energy or angular
momentum). More generally, they may be polynomial.  To discuss the linear
functions, it is helpful to notice that the {\em topological} dual space of
$\H$ can conveniently be identified with $\H$ itself using the symplectic
form: to each $Y\in\H$, we associate the linear map
$$
X\in\H\mapsto s(Y,X)\in \R.
$$
One has, from (\ref{eq:poisson}), for every $Y_1, Y_2\in\H$,
\begin{equation}\label{eq:poissonbrackets}
\{s(Y_1,\cdot),\ s(Y_2, \cdot)\} = s(Y_1, Y_2).
\end{equation}
It is then convenient to introduce
\begin{equation}\label{eq:weylclassical1}
V_\c(Y) = \e^{ -\I s(Y, \cdot)}
\end{equation}
which serves as a generating function for monomials of the type
$$
s(Y_1,\cdot) s(Y_2, \cdot) \dots s(Y_n, \cdot)=\frac{(i\partial)^n}{\partial
t_1\partial t_2 \dots\partial t_n} V_\c(t_1Y_1 +\dots t_n
Y_n)_{|t_1=0=t_2\dots=t_n}.
$$
It is immediate from the definition of the $V_c(Y)$ that
$$
V_\c(Y)\circ \Phi_t = V_\c(\Phi_{-t}Y).
$$

Working in the Schr\"odinger representation, the quantum mechanical analogues
of the $V_\c(Y)$ are the Weyl operators
\begin{equation}
V(Y)=\e^{ -\I (a\cdot P-b\cdot Q)},\ \mathrm{where}\ Y=(a,b)\in\H.
\end{equation}
The $V(Y)$ are clearly unitary operators on $L^2(\R^n)$ and satisfy the
so-called Weyl relations
$$
V(Y_1)V(Y_2)=\e^{-\frac{\I }{2} s(Y_1, Y_2)}
V(Y_1+Y_2),\qquad\forall\ Y_1, Y_2\in\H.
$$
In a  Fock representation (determined by a choice of $\Omega$), one has, with
the notation of Sect. \ref{s:focrepfin}
$$
T_\Omega V(Y) T_\Omega^{-1} =  \Wf(z_\Omega(Y)).
$$
Here the $\Wf(z_\Omega(Y))$ are the Weyl operators on the
symmetric Fock space $\eff^+(\C^n)$, as introduced in  Sect.
\ref{s:weyloperators}.

In the algebraic approach to quantum  theory, one postulates that
the interesting observables of the theory include at least those
that can be written as finite sums of $V(Y)$. One therefore
considers the algebra
$$
\mathrm{CCR}_0(\R^{2n})=\mathrm{span}\ \{ \Wf(z_\Omega(Y))\ |\ Y\in\R^{2n}\}=
\mathrm{span}\ \{ \Wf(\xi)\ |\ \xi\in\C^n\}.
$$
This algebra is irreducible. This means that the only closed
subspaces of $L^2(\R^n)\cong\eff^+(\C^n)$ invariant under the
above algebra are the trivial ones and is equivalent, via Schur's
Lemma, to the statement that the only bounded operators that
commute with all $F$ in the algebra are the multiples of the
identity. For a simple proof of these facts one may consult
\cite{db1}. This implies via a well known result in the theory of
von Neumann algebras (see \cite{br1}, for example) that its weak
closure is all of $\B(\eff^+(\C^n))$: in this sense, ``any bounded
operator on  Fock space can be approximated (in the weak
topology!) by a function of $Q$ and $P$.'' This is clearly a way
of saying that the original algebra is quite large. Note
nevertheless that its operator norm closure (called the
CCR-algebra over $\R^{2n}$ and denoted by CCR$(\R^{2n})$ is much
smaller, since it contains no compact operators. For the purposes
of these notes, I will consider CCR$(\R^{2n})$) or
CCR$_0(\R^{2n})$ as ``the'' observable algebra of the systems
considered.

Remark that  these algebras do, as sets, not depend on $\Omega$.  But again,
in close analogy to what we observed in Sect. \ref{s:focrepfin}, given an
operator on Fock space belonging to one of these algebras, its expression in
terms of $Q$ and $P$ does depend on $\Omega$, and so does therefore its
physical interpretation as an observable. So it is not only the identification
of the appropriate observable algebra which is important, but the labeling,
within this algebra, of the elements that describe the relevant physical
observables. This will be crucial once we discuss local observables in Section
\ref{ch:locobssta}, and become hopefully quite a bit clearer then too.

It is obviously not of much interest to discuss observable
algebras if one is not going to say how the observables evolve in
time. In finite dimensional systems, one is given a Hamiltonian
$H$, which is a self-adjoint operator on $L^2(\R^n)\cong
\eff^+(\C^n)$. It generates the so-called Heisenberg evolution of
each observable $F$, which is defined by $ \alpha_t(F)=\e^{\I \I
Ht} F \e^{ -\I Ht}. $ It has to be checked that the algebra of
observables and $H$ are such that this defines an automorphism of
the algebra ({\em i.e.} so that $\e^{\I Ht} F \e^{ -\I Ht}$ still
belongs to the algebra if $F$ does).

That $\alpha_t$ is an automorphism of the CCR algebra is not true in general.
 For example, it is proven in \cite{fv} that, when $H(\lambda)=\frac12
P^2 +\lambda V$, with $V$ a bounded $L^1$ function, then the
Heisenberg evolution leaves the CCR algebra invariant for all
values of $t$ and of $\lambda$ if and only if $V=0$. In other
words, the CCR algebra cannot possibly be a suitable algebra to
describe most standard quantum mechanical systems with a finite
number of degrees of freedom.

An exception to this rule are systems described by quadratic
hamiltonians, which are precisely the ones we are interested in here. An easy
example is provided by quadratic Hamiltonians of the type $H=\frac12 P^2 +
\frac12 Q\cdot \Omega^2Q$ in view of
$$
\e^{\I \d\Gamma(\Omega) t} \Wf(\xi)\e^{ -\I \d\Gamma(\Omega) t} =
\Wf(\e^{-\I \Omega t}\xi),\ \forall \xi \in \K^\C,
$$
which follows immediately from the discussion in Sect.
\ref{s:fockreprgeneral}. This clearly implies that the dynamics leaves the CCR
algebra $\mathrm{CCR} (\R^{2n})$ invariant. Note that this will work in
infinite dimensional systems just as well as in finite dimensional ones.

The discussion carries over to the infinite dimensional case without change.
One  defines the algebra of observables in the quantum theory to be
$$
\mathrm{CCR}_0(\H)= \mathrm{span}\ \{ \Wf(z_\Omega(Y))\ |\ Y\in\H\}=
\mathrm{span}\ \{ \Wf(\xi)\ |\ \xi\in\K^\C\}.
$$
Again, this algebra is independent of $\Omega$ and turns out to be irreducible
\cite{br1}, so that its weak closure is the algebra of all bounded
operators on Fock space. Its norm closure, which is much smaller, is the
so-called CCR-algebra over $\H$, for which I will write CCR$(\H)$. Since we
will only work with quadratic Hamiltonians, this
algebra is adequate for the description of such systems since it is then
invariant under the dynamics.
 Here also, to no one's surprise by now, I
hope, the interpretation of a given operator in the algebra as an observable
\emph{will} depend on $\Omega$, as we will see in more detail in Section
\ref{ch:locobssta}.

For further reference, let me define also the algebra
$$
\mathrm{CCR}_{0}(\M)=\mathrm{span}\ \{\Wf(z_\Omega(Y))|Y\in\M\},
$$
whenever $\M$ is a vector subspace of $\H$ (even if $\M$ is not
symplectic). In many situations it is natural and elegant not to
work with the norm closure of the $\mathrm{CCR}_{0}(\M)$, but with
their weak closure, for which I shall write $
\mathrm{CCR}_{\mathrm{w}}(\M)$. Further developments concerning
the CCR can be found in the contribution of J. Derezinski in this
volume \cite{d}.

%%%%%%%%%%%%%%%%%%%%%%%%%%%%%%%%%%%%%%%%%%%%%%%%%%
%%%%%%%%%%%%%%%%%%%%%%%%%%%%%%%%%%%%%%%%%%%%%%%%%%%%%
%%%%%%%%%%%%%%%%%%%%%%%%%%%%%%%%%%%%%%%%%%%%%%%%%%

\section{Local observables and local states}\label{ch:locobssta}
\subsection{Introduction}\label{s:introlocobssta}
The issue of what are local observables, local states and local
measurements has attracted a fair amount of attention and has
generated some surprises and even some controversy in the
mathematical physics literature on relativistic quantum field
theory. The controversy has centered on the question of particle
localization, of possible causality violations and of relativistic
invariance. I will address these issues in  the present section
within the restricted context of the
 free oscillator fields under study here, some of which are relativistically
 invariant, while others are not. I will  argue that there is
 not much reason to be surprised and certainly no ground for controversy.

 After defining what is meant by a local observable (Sect. \ref{s:deflocstr})
 and giving some examples (Sect. \ref{s:exalocstr}),
  the notion of ``strictly local excitation of the vacuum'' is introduced in Sect. \ref{s:strlocvacexc}. I will then state a generalization of a theorem of Knight asserting that, if $\Omega$ is a non-local operator, then states with a finite number of field excitations
 cannot be strictly local excitations of the vacuum (Sect. \ref{s:knitherev}).
 It will be shown  through examples (Sect. \ref{s:trainvsys})
 that the above condition on $\Omega$ is typically satisfied in models of
 interest and I will explain the link between the above notion of
 localized excitation of the vacuum
 and the so-called Newton-Wigner localization (Sect. \ref{s:newwig}).
 It will be argued that the latter is not
 a suitable notion to discuss the local properties of the states of oscillator fields. The actual proof of Knight's theorem is deferred to Sect. \ref{s:prothelocalquanta}.

%%%%%%%%%%%%%%%%%%%%%%%%%%%%%%%%%%%%%%%%%%%%%%%%%
%%%%%%%%%%%%%%%%%%%%%%%%%%%%%%%%%%%%%%%%%%%%%%%%%%%
%%%%%%%%%%%%%%%%%%%%%%%%%%%%%%%%%%%%%%%%%%%%%%%

\subsection{Definition of a local structure}\label{s:deflocstr}
Among the interesting observables of the oscillator systems we are
studying are certainly the ``local'' ones. I will give a precise
definition in a moment, but thinking again of the oscillator
chain, ``the displacement $q_7$ of the seventh oscillator'' is
certainly a ``local'' observable. In the same way, if dealing with
a wave equation, ``the value $q(x)$ of the field at $x$ '' is a
local observable. The Hamiltonian is on the other hand not a local
observable, since it involves sums or integrals over all
oscillator displacements and momenta. Generally, ``local
observables'' are functions of the fields and conjugate fields in
a bounded region of space. Of course, this notion does not make
sense for all harmonic systems, defined by giving a positive
operator $\Omega^2$ on some abstract Hilbert space $\K$. So let me
reduce the level of abstractness of the discussion, therefore
hopefully increasing its level of pertinence, and define what I
mean by a system with a local structure.

In view of what precedes, I will limit my attention to  free
oscillator fields over a real Hilbert space $\K$ of the form $\K =
\L2r(K, \d \mu)$, where $K$ is a topological space and $\mu$ a
Borel measure on $K$. Here the subscript ``$\R$'' indicates that
we are dealing with the real Hilbert space of real-valued
functions. In fact, all examples I have given so far are of the
above type.
\begin{definition} \label{def:local} A local structure for the oscillator field determined by
$\Omega$ and $\K = \L2r(K, \d \mu)$ is a subspace $\S$ of $\K$ with
the following properties:
\begin{enumerate}
\item $\S\subset \K_{1/2}\cap\K_{-1/2}$; \\
\item Let $B$ be a Borel subset of $K$, then $\S_B\equiv\S\cap
\L2r(B, \d \mu)$ is dense in $\L2r(B, \d \mu)$.
\end{enumerate}
\end{definition}
This is a pretty strange definition, and I will give some examples in a
second, but let me first show how to use this definition to define what is
meant by ``local observables''. Note that, thanks to the density condition
above,
$$
\H(B,\Omega)\stackrel{\mathrm{def}}{=}\S_B\times \S_B
$$
is a symplectic subspace of $\H$ so that the restriction of $\Wf\circ
z_\Omega$ to $\H(B,\Omega)$ is a representation of the CCR over
$\H(B,\Omega)$.
\begin{definition} \label{def:localobs}
Let $\K= \L2r(K, \d \mu), \Omega, \S$ be  as above and let $B$ be
a Borel subset of $K$. The algebra of local observables over $B$
is the algebra
$$
\mathrm{CCR}_0(\H(B,\Omega))=\mathrm{span}\ \{\Wf(z_\Omega(Y))\ | \
Y\in\S_B\times\S_B\}.
$$
\end{definition}
Note that  $\Omega$ plays a role in the definition of $\S$ through
the appearance of the spaces $\K_\lambda$. The first condition on
$\S$ guarantees that $\S\times\S\subset \H$ so that, in
particular, for all $Y\in\S\times\S$, $s(Y,\cdot)$ is well defined
as a function on $\H$ which is  important for the definition of
the local observables to make sense. In practice, one wants to be
able to use the same spatial structure $\S$ for various choices of
$\Omega$, in order to be able to compare different systems built
over the same space $\K=\L2r(K, \d \mu)$. Note nevertheless that
even then, the algebras
 of local and of quasi-local
observables, which are algebras of bounded operators on the Fock space
$\eff^+(\K^\C)$ do, \emph{as sets}, depend on $\Omega$. This is in contrast to
the algebra  of ``all'' observables,
$$
\mathrm{CCR}_0(\H)= \mathrm{span}\ \{\Wf(z_\Omega(Y))\ | \ Y\in\H\} =
\mathrm{span}\ \{\Wf(\xi)\ | \ \xi\in\Kc\},
$$
which is, as a set, independent of $\Omega$, as pointed out before. In other
words, some of the physics is hidden in the way the local algebras are
imbedded in the CCR algebra over $\H$.

%%%%%%%%%%%%%%%%%%%%%%%%%%%%%%%%%%%%%%%%%%%%%%%%%%%%%%%%%%%
%%%%%%%%%%%%%%%%%%%%%%%%%%%%%%%%%%%%%%%%%%%%%%%%%%%%%%%%%%%%%

\subsection{Examples of local structures}\label{s:exalocstr}
\subsubsection{Oscillator lattices -- Klein-Gordon equations}
In the case of the  translationally invariant oscillator lattices in dimension
$2$ or higher presented in Sect. \ref{s:examples2}, $\S$ can be taken to be
the space of sequences $q$ of finite support, even in the massless case, as is
easily checked. Alternatively, you could take $\S$ to be the larger space of
sequences of fast decrease. This has the advantage that then $\S\times \S$ is
dynamics invariant. Note that in neither of these examples
 $\S\times \S$ is  $J$ invariant, though, so that $\S\times \S$ will
not be a complex vector subspace of $(\H, J)$, just a real one. This is also
true for $\S_B\times \S_B$ and will be crucial when discussing ``local
excitations of the vacuum'' in quantum field theory.

\begin{exercise} Check all of the above statements in detail.
\end{exercise}

As an example of a local observable, we have, with $\eta\in\S$ of bounded
support in some set $B\subset \Z^d$,
$$
\e^{\I \eta \cdot Q} = \Wf(\frac{\I }{\sqrt2}\Omega^{-1/2}\eta).
$$
Very explicitly, one may think of taking $\eta(j)=\delta_{j,k}$ and then this
is
 $\e^{\I  Q(k)}$, a simple function of the displacement of the oscillator at site $k\in\Z^d$. At the risk of boring the wits out of you, let me point out yet again that this \emph{fixed observable} is represented on Fock space by a \emph{different operator} for different choices of $\Omega$.

Similarly
$$
\e^{\I \eta \cdot P} = \Wf(-\frac{1}{\sqrt2}\Omega^{1/2}\eta)
$$
is a function of the momenta of the oscillators in the support of $\eta$.

In the one-dimensional translationally invariant lattice a spatial structure
does not exist when $\nu=1/2$ because of the strong infrared singularity.
Indeed, due to the density condition in the definition of the local structure,
it is clear that $\S$ must contain all sequences of finite support, and those
do not belong to $\K_{-1/2}$ in dimension $1$, as we already pointed out in
Sect. \ref{s:examples2}.

%\subsection{Wave and Klein-Gordon equations}

Similarly, the wave and Klein-Gordon equations on $\R^d$ admit for example
$C_0(\R^d)$ or the space of Schwartz functions as a spatial structure in
dimension $2$ or higher, as follows from the discussion in Sect.
\ref{s:examples3}.

\subsubsection{The finite dimensional case}\label{s:findimcas}
I find this example personally most instructive. It forces one into an unusual
point of view on a system of $n$ coupled oscillators that is well suited to
the infinite dimensional case. Think therefore of a system of $n$ oscillators
characterized by a positive $n$ by $n$ matrix $\Omega^2$, as in Sect.
\ref{s:examples1}. A local observable of such a system should be a function of
the positions and momenta of a fixed finite set of oscillators. Does the
definition given above correctly incorporate this intuition? Let's check.

In this case, $\K=\R^n$, which I view as $\L2r(K)$, where $K$ is simply the
set of $n$ elements. Indeed, $q\in\R^n$ can be seen as a function
$q:j\in\{1,\dots n\}\mapsto q(j)\in\R$, obviously square integrable for the
counting measure. I already explained in detail the identification between the
quantum state space $L^2(\R^n)$ and $\eff^+(\C^n)$ (Sect.
\ref{s:focrepfin}). Here $\C^n$ is the complexification of $\R^n$, and as such
naturally identified with $L^2(K, \C)$. So, finally
$$
L^2(\R^n)\cong \eff^+(L^2(K, \C)).
$$
Consider now a subset $B$  of $K$, say $B=\{1, 6, 9\} (n\geq 9)$. It is an
excellent exercise to convince oneself that, unraveling the various
identifications, a local observable over $B$ is a finite linear combination of
operators on $L^2(\R^n)$ of the form $(a_j, b_j\in\R, j\in B)$:
$$
\exp{ -\I \left(\sum_{j\in B} (a_jP_j-b_jQ_j)\right)}.
$$
Better yet, if you write (with $\sharp B$ denoting the cardinality
of the set $B$)
$$
L^2(\R^n)\cong L^2(\R^{\sharp B}, \prod_{j\in B}\d x_j)\otimes
L^2(\R^{n-\sharp B}\prod_{j\not\in B}\d x_j),
$$
then it is clear that the weak closure of the above algebra is
$$
{\cal B}(L^2(\R^{\sharp B}, \prod_{j\in B}\d x_j))\otimes \bbbone.
$$
So, indeed, a local observable is clearly one that involves only the degrees
of freedom indexed by elements of $B$.

\begin{exercise} Convince yourself all of this is true.
\end{exercise}

%%%%%%%%%%%%%%%%%%%%%%%%%%%%%%%%%%%%%%%
%%%%%%%%%%%%%%%%%%%%%%%%%%%%%%%%%%%%%%%
\subsubsection{Unbounded local observables}\label{s:unblocobs}
To make contact with the physics literature, it will be convenient
on occasion in the following to refer to polynomials in
$\frac{\d}{\d t}W(z_\Omega(tY))|_{t=0}$ with $Y\in\H(B,\Omega)$
as local observables over $B$ as well. These are sums of
expressions of the form
$$
\Pi_{S}(z_\Omega(Y_1))\Pi_{S}(z_\Omega(Y_2))\dots\Pi_{S}(z_\Omega(Y_n))
$$
where each $Y_j\in\H(B,\Omega)$ Alternatively and perhaps more suggestively,
these are sums of expressions of the form
$$
(\eta_1\cdot Q)\dots(\eta_m\cdot Q)\quad \mathrm{and}\quad (\eta_1\cdot
P)\dots(\eta_m\cdot P),
$$
or of products thereof, where each $\eta_j\in\S_B$. Again, for lattices, these
are polynomials in the positions and momenta of the individual oscillators in
some subset $B$ of the lattice $\Z^d$.

%%%%%%%%%%%%%%%%%%%%%%%%%%%%%%%%%%%%%%%
%%%%%%%%%%%%%%%%%%%%%%%%%%%%%%%%%%%%%%%
\subsection{Strictly localized vacuum excitations}\label{s:strlocvacexc}
I now want to give meaning to the notion of ``local excitation of
the vacuum'' for  general free oscillator fields with a local
structure $\S$. So in this section $\K=L_{\mathrm r}^2(K,\d \mu)$,
and $\S$ satisfies the conditions of Definition \ref{def:local}.

The equivalent classical notion is readily described and was already discussed
in Sect. \ref{s:examples2}.  The vacuum, being the ground state of the
system, is the quantum mechanical equivalent of the global equilibrium $X=0$,
which belongs of course to the phase space $\H$, and a local perturbation of
this equilibrium is an initial condition $X=(q,p)\in\S\times \S$ with the
support of $q$ and of $p$ contained in a (typically bounded) subset $B$ of
$K$. An example of a local perturbation of an oscillator lattice is a state
$X\in\H$ where only $q_0$ and $p_0$ differ from $0$. In the classical theory,
local perturbations of the equilibrium are therefore states that differ from
the equilibrium state only inside a bounded subset $B$ of $K$. It is this last
formulation that is readily adapted to the quantum context, through the use of
the notion of ``local observable'' introduced previously.

For that purpose, we first introduce the following notion, which is due to
Knight \cite{kn}.
\begin{definition}\label{def:indistinguishable} Let $\psi,\psi'\in\eff^+(\Kc)$.
We will say that $\psi$ and $\psi'$ are indistinguishable inside a Borel set
$B\subset K$ if, for all $X\in\H(B,\Omega)$,
\begin{equation}\label{eq:indistinguishable}
\langle\psi|\Wf(z_\Omega(X))|\psi\rangle =
\langle\psi'|\Wf(z_\Omega(X))|\psi'\rangle.
\end{equation}
\end{definition}
Note that, given $\psi$ and $B$, it is easy to construct many states that are
locally indistinguishable from $\psi$ in $B$. Indeed, one may consider
$\Wf(z_\Omega(X))|\psi\rangle$, for any $X\in\H({B^c}, \Omega)$.

We are now ready to define what we mean by a strictly local excitation of
the vacuum.
\begin{definition} \label{def:locexc} If $B$ is a Borel subset of $K$, a strictly
local excitation of the vacuum with support in $B$ is a normalized
vector $\psi\in\eff^+(\Kc)$, different from the vacuum itself,
which is indistinguishable from the vacuum outside of $B$. In
other words,
\begin{equation}\label{eq:localexcitation}
\langle \psi | \Wf(z_\Omega(Y))|\psi\rangle =  \langle 0 |
\Wf(z_\Omega(Y))|0\rangle
\end{equation}
for all $Y=(q,p)\in\H({B^c,\Omega)}$.
\end{definition}

For brevity, I will occasionally call such states ``local
states'', although this terminology conjures up images that are
misleading. In view of what precedes, the coherent states
$\Wf(z_\Omega(X))|0\rangle$, for any $X\in \H(B,\Omega)$ are
strictly local excitations of the vacuum in $B$. The use of the
adjective ``strictly'' is motivated by the possibility of relaxing
condition (\ref{eq:localexcitation}) to allow for states that are
only approximately localized in $B$, but for which the expectation
values of observables located far from $B$ converge more or less
rapidly to the corresponding vacuum expectation values. I refer to
\cite{db2} for details.

%This issue is
%discussed in Sect. \ref{s:loctai}.

%%%%%%%%%%%%%%%%%%%%%%%%%%%%%%%%%%%%%%%%%%
%%%%%%%%%%%%%%%%%%%%%%%%%%%%%%%%%%%%%%%%%%
\subsection{Knight's theorem revisited}\label{s:knitherev}
Recall that states with a finite number of field quanta, {\em
i.e.}  states belonging to $\eff^{\mathrm{fin},+}(\K^\C)$, are
interpreted as states describing a finite number of
quasi-particles (see Sect. \ref{s:parint}). Hence one natural
question is whether such a state can be a strictly local
excitation of the vacuum in a set $B$. Theorem
\ref{thm:localquanta} below gives a necessary and sufficient
condition for this to happen.

First, I need a definition:
\begin{definition}\label{def:locop}
$\Omega$ is said to be strongly non-local on $B$ if there does not exist a
non-vanishing $h\in\K_{1/2}$ with the property that both $h$ and $\Omega h$
vanish outside $B$.
\end{definition}
Here I used the further definition:
\begin{definition}\label{def:supph} Let $h\in\K_{\pm1/2}$ and $B\subset K$. Then $h$ is said to vanish in $B$ if for all $\eta\in\S_{B}$, $\eta\cdot h=0$. Similarly, it is said to vanish outside $B$, if for all
$\eta\in\S_{B^{\mathrm c}}$, $\eta\cdot h=0$.
\end{definition}
Note that this definition uses the density of $S_B$ in $L^2(B)$ implicitly,
because without this property, it  would not make much sense.
 Intuitively, a strongly non-local operator is one
that does not leave the support of any function $h$ invariant.

\begin{theorem} \label{thm:localquanta} Let $B$ be a Borel subset of $K$.
Then the following are equivalent:

(i) $\Omega$ is strongly non-local on $B$;

(ii) There do {\em not} exist   states in $\eff^{{fin},+}(\K^\C)$
which are strictly strictly local excitations of the vacuum with
support in $B\subset K$;
\end{theorem}

I will give the proof of this result in Sect. \ref{s:prothelocalquanta}.

 Statement (i) of the theorem gives a more or less easily checked neccessary and sufficient condition for the non-existence of
states with a finite number of field quanta that are localized in a  region
$B$. I will show in the examples developed in the following sections
 that this condition is so to speak always satisfied when $B$ is a bounded set: I mean,
it is satisfied in the various models that are typically studied
in solid state physics, in relativistic quantum field theory, or
in the theory of free quantum fields on curved space-times.
Indeed, in these examples, $\Omega^2$ is a finite difference or
(second order elliptic) differential operator, so that it is
local: it preserves the support. But its positive square root,
$\Omega$, is more like a pseudo-differential operator, and
therefore does not preserve supports. This will be shown in
several cases below. The upshot is that states with a finite
number of particles, and a fortiori, one-particle states, are
never strictly localized in a bounded set $B$. This gives a
precise sense in which the elementary excitations of the vacuum in
a bosonic field theory (relativistic or not) differ from the
ordinary point particles of non-relativistic mechanics: their
Hilbert space of states contains no states in which they are
perfectly localized.

So, to sum it all up, you could put it this way. To the question
\begin{quote}
{Why is there no sharp position observable for  particles?}
\end{quote}
the answer is
\begin{quote}{It is the non-locality of $\Omega$, stupid!}
\end{quote}

Should all this make you feel uncomfortable, I hope the further discussion in
Sects. \ref{s:newwig} of the history of the quest for a
``position observable'' in relativistic field theory will be of some help.

\subsection{Examples}\label{s:trainvsys}
As a warm-up, here is my favourite example.

\begin{exercise} Let $\K=\R^2$ so that $\Omega^2$ is a two by two matrix and,
as explained in Sect. \ref{s:findimcas}, $K=\{1,2\}$. Show that in this
case, a state with a finite number of quanta can be a strictly localized
excitation of the vacuum on $B=\{1\}$ only if $\Omega^2$ is diagonal. In other
words, this can happen only if the two oscillators are not coupled.
\end{exercise}

For typical translationally invariant systems, it is easy to see $\Omega$ is
strongly non-local over bounded sets, so that we can conclude there are no
strictly localized finite particle states. This is the content of the
following results.

\begin{theorem} Let $\K= \L2r(\R^d, \d x)$  and let
$\omega$ be a positive function belonging to
$L^\infty_{\mathrm{loc}}(\R^d, \d k)$ with $\omega^{-1}\in
L^1_{\mathrm{loc}}(\R^d, \d k)$. Suppose both $\omega$ and
$\omega^{-1}$ are polynomially bounded at infinity.  Let
$\Omega=\omega(|\nabla|)$. Then $\S=\S(\R^d)$ is a local structure
for this system. If $\omega$ does not extend to a holomorphic
function on the complex plane, then $\Omega$ is strongly non-local
on any bounded open set $B$. Consequently, there exist no states
with a finite number of quasi-\-particles that are strictly
localized excitations of the vacuum in such a set $B$.
\end{theorem}

The proof is a simple application of the Paley-Wiener theorem together with
Theorem \ref{thm:localquanta}. Note that the theorem applies to the
Klein-Gordon equation: so we recover in this way Knight's original result.
Pushing the use of the Paley-Wiener theorem a little further, one can also
prove:

\begin{theorem} Let $\K= \L2r(\R^d, \d x)$  and $\Omega^2=-\Delta+m^2$,
with $d\geq1, m>0$, or $d\geq 2, m\geq0$. Then $\S=\S(\R^d)$ is a local
structure for this system and there exist no states with a finite number of
quasi-\-particles that are strictly localized excitations of the vacuum in any
set $B$ with non-empty open complement.
\end{theorem}

 The result one needs here is proven in \cite{sego}: for
 $\Omega=\sqrt{-\Delta +m^2}$,
 $h$ and $\Omega h$ cannot both vanish on the same open set.
 Via  Theorem \ref{thm:localquanta} this implies the above result.

An analogous result holds for the translationally invariant lattices discussed
in Sect. \ref{s:examples2}. In particular,  with $\Omega^2$ as in
(\ref{eq:lattice}), it is very easy to see that there are no states with a
finite number of quanta that are perfectly localized perturbations of the
vacuum on a finite number of lattice sites. The spatial structure is given
here by the sequences of finite support, as discussed in Sect.
\ref{s:exalocstr}.

Similarly, for the wave and Klein-Gordon equations the operator $\Omega$ is typically
also strictly non-local, but I will not go into this here.

It is clear from these examples that Knight's theorem has less to
do with relativity than with coupled oscillators, which is the
point I wanted to make all along.

%%%%%%%%%%%%%%%%%%%%%%%%%%%%%%%%%%%%%%%
%%%%%%%%%%%%%%%%%%%%%%%%%%%%%%%%%%%%%%
%%%%%%%%%%%%%%%%%%%%%%%%%%%%%%%%%%%%%%

\subsection{Newton-Wigner localization}\label{s:newwig}
Knight's result appears  counterintuitive. Indeed, we argued first
that the Fock space structure of the Hilbert space of states of
the field invites a particle interpretation (Sect.
\ref{s:parint}), we then introduced what looks like a perfectly
reasonable notion of ``strictly localized excitation of the
vacuum'',  only to end up discovering that states with a finite
number of particles cannot be strictly localized. Since the notion
of a particle evokes an entity that is localized in space, this
may seem paradoxical. My point of view is simple:  the way out of
this paradox is, as I have suggested before (Sect.
\ref{s:parint}), that one has to keep in mind that the particles
under discussion here are just {\em excited states of an extended
system} and that, just like in an oscillator ring, chain, or
lattice, the analogy with the point particles of elementary
classical or quantum mechanics courses should not be pushed too
far. Calling those excitations  particles amounts to nothing more
than an occasionally confusing abuse of language. The lesson to be
learned from Knight's result is therefore that such field quanta
may carry momentum and energy, but they cannot be perfectly
localized. Viewed from the angle I have chosen, this is not even
surprising. The examples showed indeed this statement is true in a
system with two oscillators, and in oscillator lattices. One
should in particular not hope to associate a position operator
with those quanta, having all the usual properties familiar from
the description of point particles in ordinary Schr\"odinger
quantum mechanics.

I could end the story there. But a very different point of view,
based precisely on the use of a position operator (the so-called
Newton-Wigner position operator) to locate the particles, was
developed well before Knight's work, in the context of (free)
relativistic quantum field theory of which the Klein-Gordon field
is a particular example. Since this alternative point of view has
met with a certain amount of popularity, it cannot be dismissed
too lightly. Below I will explain it has an obvious analog for the
oscillator systems under study here and I will show why, although
it seems at first sight perfectly natural, it is clearly
ill-conceived. The implication of this remark for the debate about
supposed causality problems in relativistic quantum field theory
and a further overview of some other  issues related to ``particle
localization'' in that context  will also be given.

%%%%%%%%%%%%%%%%%%%%%%%%%%%%%%%%%%%%%%%%%%%
\subsubsection{Newton-Wigner localization: the definition}\label{s:newwiga}
Let us therefore turn again to an oscillator field with spatial
structure so that $\K^\C=L^2(K, \d \mu, \C)$. The state space of
this system is the bosonic Fock space $\eff^+(\Kc)$ of which
$\K^\C=L^2(K, \d \mu, \C)$ represents the one-particle sector.
Now, if the system is in the state $\psi\in\Kc\subset
\eff^+(\Kc)$, it is in view of the particle interpretation of the
field states explained in Sect. \ref{s:parint} very tempting to
interpret $\mid\psi\mid^2(y)\d \mu$ as the probability for finding
the ``particle'' in a volume $\d \mu$ around $y$, or in a preciser
manner, to say that the probability for finding the particle in
$B\subset K$ is given by
$$
\int_B\mid\psi\mid^2(y)\d \mu.
$$
This seems like a quite reasonable thing to do because it is
completely analogous to what is done in the non-relativistic
Schr\"odinger quantum mechanics of particle systems. I will call
the projection valued measure $B\mapsto \chi_B$, where $\chi_B$ is
the operator of multiplication by the characteristic function of
$B$ the Newton-Wigner position observable.  If $\psi \in L^2(K, \d
\mu, \C)$ is supported in $B\subset K$, we say $\psi$ is
``Newton-Wigner localized in $B$''. This terminology is inspired
by the observation that, when considering the particular example
of an oscillator field given by the wave or Klein-Gordon equation,
one has $\K=L^2(\R^3, \d x, \C)$ and in that case the above
measure is indeed the joint spectral measure of the usual
Newton-Wigner position operator of relativistic quantum field
theory \cite{nw}. This choice of position observable may seem
reasonable, but it is only based on an analogy, and as I will now
show, it is not reasonable at all.

For that purpose, let us go back to  the particular example of the
oscillator ring treated before (Sects. \ref{s:examples1} and
\ref{s:findimcas}) and see what the Newton-Wigner position
operator means in that case.  Remember, this is just a system of
$n$ coupled oscillators. So the quantum Hilbert space can on the
one hand be seen as $L^2(\R^n, \d x)$ (Schr\"odinger
representation) and the system can be studied through the
displacements and momenta of those oscillators. This is the usual
point of view. Alternatively, it be identified with the bosonic
Fock space $\eff^+(\Kc)$ (Fock representation), where now the
one-particle subspace is $\Kc=\C^n$. The latter, as explained in
Sect. \ref{s:findimcas}, can be thought  of as $L^2(\Z/n\Z, \C)$.
In other words,  it is tempting to interpret $\psi\in\Kc$ as the
quantum mechanical state of a ``particle'' hopping along $n$
sites! Its probability of being at site $i$ is then given by
$|\psi(i)|^2$. More generally, any state of the $n$ oscillators
can be seen as a superposition of $0,1,2,\dots$ ``particle''
states, where now ``particle'' refers to an imagined entity
hopping along the sites of the chain.  Speaking like this, we are
pushing the particle interpretation maximally.  The state
$a^\dagger(\delta_i)|0\rangle$ is then thought of as particle
perfectly localized on the site $i$.

But does this make sense? Certainly, whatever picture used, the
mean square displacement of the oscillator at site $j$ is a
relevant physical observable in this system. The problem is that
this mean square displacement will differ from its vacuum value if
$j\not= i$:
$$
\langle0|a(\delta_i)Q_j^2a^\dagger(\delta_i)|0\rangle \not=
\langle0|Q_j^2|0\rangle.
$$
So the idea that the system contains only one particle, and that the latter is
localized perfectly at $i$, the rest of the sites being ``empty'', is not
tenable. Indeed, if the particle is at site $i$, and if space (here
represented by the $n$ sites) is otherwise ``empty'', how can any observable
at site $j$ take a value different from its vacuum value? The problem is of
course readily solved if one stops trying to interpret the quantity
$|\psi(i)|^2$ as a probability of presence for a particle.

The same analysis carries immediately over to the oscillator
chains or lattices discussed before. It is perhaps even more
telling there. Now the one particle space is $\ell^2(\Z^d, \C)$
and so the idea of thinking of states in this space as describing
a particle hopping on the sites of the lattice $\Z^d$ may seem
even more reasonable. Models of this type are used in solid state
physics to describe lattice vibrations, and the quanta are then
called phonons.  They are excitations of the oscillator lattice
and -- as Knight's theorem tells us -- cannot be perfectly
localized in the sense that, if the system is in a one-phonon
state $\psi\in \ell^2(\Z^d)$, then it cannot coincide with the
vacuum outside a finite subset of the lattice. This does not lead
to any interpretational difficulties, {\em as long as one does not
try to interpret} $|\psi(i)|^2$  {\em as the probability of
finding the particle at site} $i$ {\em of the lattice}.

Finally, without any change whatsoever, the same analysis carries
over to the Klein-Gordon equation. Let $B$ be a bounded subset of
$\R^3$ and $\psi\in L^2(\R^3, \d x, \C)$ be supported in $B$. As
Knight's theorem tells us, the corresponding one-particle state of
the field is not an excitation of the vacuum localized inside $B$.

The conclusion I draw from all this is that {\em the Newton-Wigner
operator does not provide an appropriate tool  to describe the
strict localization properties of the states of extended systems
of the type discussed here}. It shows up only because of  an
understandable but ill-fated desire to force a {\em particle
interpretation} with all its usual attributes on the states of a
{\em field}. The right notion of a (strictly) localized state is
the one given by Knight (Definition \ref{def:locexc}). This has
lead to some debate in the context of relativistic quantum field
theory, upon which I shall comment below. Anticipating on the
discussion there, I would like to stress that my line of argument
here, and in particular my criticism of the use of the
Newton-Wigner operator has nothing to do with relativity, or with
causality, but is related instead to the fact that we are dealing
with extended systems.

%%%%%%%%%%%%%%%%%%%%%%%%%%%%%%%%%%%%%%%%%%%%%%%%%%%%%%%%%
%%%%%%%%%%%%%%%%%%%%%%%%%%%%%%%%%%%%%%%%%%%%%%%%%%%%%
%%%%%%%%%%%%%%%%%%%%%%%%%%%%%%%%%%%%%%%%%%%%%%%%%%%%%%%%%%

\subsubsection{Causality problems}\label{s:newwigb}
In the early days of relativistic quantum field theory, and well before
anything like Knight's theorem was formulated or proven, the particle
interpretation of the field states made it perfectly natural to search  for  a
position operator with the usual properties familiar from non-relativistic
quantum mechanics. In other words, if the field quanta are particles, one
would want to answer the question: ``Where is the particle?'' It should
therefore not come as a  surprise that a fair amount of  literature was
devoted to this problem. The theory received its definite form
 in \cite{nw} and a slightly more rigorous treatment was subsequently given in \cite{wi}.
 References to earlier work can be found in those two papers and in \cite{scwi}.
 The discussion in \cite{nw} centers on the
question how to identify, inside a relativistic elementary system ({\em i.e.}
inside a unitary irreducible representation of the Poincar\'e group), a
``position operator $\hat x = (\hat x_1, \hat x_2, \hat x_3)$'', using only
natural requirements -- formulated as axioms -- on the transformation
properties of this operator under rotations and translations. The upshot of
this analysis is that such an operator exists (for most values of spin and
mass) and that it is unique. It is called the Newton-Wigner position operator
in the literature. As an example, there exists such an operator in the one
field quantum sector of the quantized Klein-Gordon field, which carries an
irreducible representation of the Poincar\'e group of zero spin and it is
precisely the one discussed in the previous subsection. Now, the joint
spectral measure of the three components of $\hat x$ defines a projection
valued measure $P_B$, where $B$ is a Borel subset of $\R^3$. If the
interpretation of $\hat x$ as a position operator along the lines of the usual
interpretational rules of quantum mechanics is to make sense, then eigenstates
of $P_B$ with eigenvalue $1$ are to be thought of as states ``perfectly
localized inside $B$''. This is referred to as NW-localization. This is
precisely the interpretation given to the Newton-Wigner operator in the
literature which is, as explained before, at odds with Knight's notion of
local excitation of the vacuum. Nevertheless, the axiomatic derivation of the
Newton-Wigner operator, and its perfect analogy with the familiar situation in
the quantum mechanics of non-relativistic particles gives it something very
compelling, which probably explains its success. As a result, some authors
have written that the Newton-Wigner operator is the only possible position
operator for relativistic quantum particles. In \cite{wi}, one reads the
following claim: ``I venture to say that any notion of localizability in
three-dimensional space which does not satisfy [the axioms] will represent a
radical departure from present physical ideas.'' Newton and Wigner say
something similar, but do not put it so forcefully: ``It seems to us that the
above postulates are a reasonable expression for the localization of the
system to the extent that one would naturally call a system unlocalizable if
it should prove to be impossible to satisfy these requirements. In \cite{scwi}
one can read: ``One either accepts the Newton-Wigner position operator when it
exists, or abandons his axioms. We believe the first alternative is well worth
investigation and adopt it here.'' I of course have argued above that one
should abandon it, and that this  neither constitutes a departure from
standard physical ideas, nor means that one abandons the notion of
localizability.

Still, even among those that have advocated the use of
Newton-Wigner localization, this notion has stirred up a fair
amount of debate, since it violates causality, as I now briefly
explain.

Indeed, first of all, a one-particle state of the Klein-Gordon field perfectly
NW-localized in some bounded set $B$ at an initial time, is easily seen to
have a non-zero probability to be found arbitrarily far away from $B$, at any
later time, violating causality. Since the theory is supposed to be
relativistic, this is a real problem that has received much attention.
Actually, replacing the projection operators $P_B$ of the NW-position operator
by any other positive operators transforming correctly under space
translations, Hegerfeldt proved that the causality problem remains (see
\cite{he1} \cite{he2} and for a more recent overview, \cite{he3}). In
addition, and directly linked to the previous observation, a state perfectly
localized in one Lorentz frame is not in another one.  These difficulties,
while well known and widely stressed, are often dismissed with a vague appeal
to one of the following  somewhat related ideas. Although the Newton-Wigner
derivation does not refer to any underlying field theory, these arguments all
involve remembering that the ``particles'' in relativistic field theory are
excitations of the field.

The first such argument goes as follows. In a field theory a
position measurement of a particle would lead to pair creation
(see \cite{scwi}) and so the appearance of particles far away is
not paradoxical. This line of reasoning is not very satisfactory
(as already pointed out in \cite{nw}), since it seems to appeal to
a (non-specified) theory of interacting fields to deal with the a
priori simple non-interacting field. An alternative argument
stresses that in a theory which allows for multi-particle states,
the observation of exactly one particle inside a bounded set $B$
entails the observation of the absence of particles everywhere
else, and is therefore not really a local measurement. As such,
the appearance later on of particles far away does not violate
causality (see \cite{fu}). This argument is certainly correct. But
it is again qualitative and nothing guarantees that it can
correctly account for the ``amount'' of causality violation
generated by the Newton-Wigner position.

All in all, it seems considerably simpler to adopt the notion of
``strictly localized vacuum excitation'' introduced by Knight,
which is perfectly adapted to the study of the extended systems
under consideration here and to accept once and for all that the
particles of field theory are elementary excitations of the system
(or field quanta) that do not have all the usual attributes of the
point particles of our first mechanics and quantum mechanics
courses. This seems to be the point of view implicitly prevalent
among physicists, although it is never clearly spelled out in the
theoretical physics textbooks for example, as I will discuss in more detail in \cite{db2}.
It also has the advantage that no causality problems arise.
Although traces of this argument can occasionally be found in the
more mathematically oriented literature, Knight's definition of a
strictly local excitation of the vacuum and his result on the
non-localizability of finite particle states seem to be mostly
ignored in discussions of the issue of the localizability of
particles in field theory, of which there continue to be many
\cite{ba} \cite{ha2} \cite{tel} \cite{stre} \cite{fl} \cite{flbu}

Having advocated Knight's definition of ``local state'', it remains to prove the
extension of his theorem given above.

%%%%%%%%%%%%%%%%%%%%%%%%%%%%%%%%%%%%%%%
%%%%%%%%%%%%%%%%%%%%%%%%%%%%%%%%%%%%%%%%%%%%%%%%%%%%

\subsection{Proof of Theorem \ref{thm:localquanta}}\label{s:prothelocalquanta}
The theorem is reduced to abstract nonsense through the following proposition.
Note that, for any subset $\M$ of a Hilbert space $\V$, $\M^{\perp}$ denotes
its orthogonal complement, which is a complex subspace of $\V$.

\begin{proposition} \label{prop:reduce} Let $\K=\L2r(K, \d \mu)$, $\Omega$ and $\S$ be as before. Let $B\subset K$. Then the following statements are equivalent.

(i) $\Omega$ is strongly non-local over $B$.

(ii) $\left(z_\Omega(\H({B^c},\Omega))\right)^\perp=\{0\}$ or, equivalently,
\begin{equation}\label{eq:condition}
\left(\overline{\mathrm{span}}_\C
z_\Omega(\H({B^c},\Omega))\right)^\perp=\{0\}.
\end{equation}
\end{proposition}

Indeed,  that Theorem
\ref{thm:localquanta} (i) and (ii) are equivalent now follows from  Theorem
\ref{thm:indistinguishable} below.

\proof (Proposition \ref{prop:reduce}) It is easy to see that $\xi\in
\left(z_\Omega(\H(B^c,\Omega))\right)^\perp$ if and only if
$$
\overline \xi\cdot \Omega^{1/2}\eta = 0 = \overline \xi\cdot
\Omega^{-1/2}\eta,
$$
for all $\eta\in \S_{B^c}$. We can suppose without loss of generality that
$\xi$ is real. Now, if $\xi\in\K$, then $\Omega^{1/2}\xi\in\K_{-1/2}$ and hence
$$
0= \xi\cdot \Omega^{1/2}\eta={\Omega^{1/2}\xi} \cdot \eta
$$
which proves $\Omega^{1/2}\xi$ vanishes outside $B$. Similarly
$\Omega^{-1/2}\xi$ vanishes outside $B$. Setting $h=\Omega^{-1/2}\xi$ the
result follows. \endproof

\begin{theorem} \label{thm:indistinguishable} Let $\W$ be a real subspace of $\V$.

(i)  If $\psi\in \eff^+((\overline{\mathrm{span}}_\C \W)^\perp)
\subset\eff^+(\V)$, $\parallel \psi\parallel=1$, then
\begin{equation}\label{eq:locvac}
\langle\psi|\Wf(\xi)|\psi\rangle =
\langle0|\Wf(\xi)|0\rangle,\qquad\forall\xi\in\W.
\end{equation}

(ii) If $\mathrm{span}_\C \W$ is dense in $\V$ then there exist no
$\psi\in\eff^{\mathrm{fin},+}(\V)$ other than $|0\rangle$ itself
so that (\ref{eq:locvac}) holds.
\end{theorem}

Clearly, the equivalence of (i) and (ii) in Theorem \ref{thm:localquanta} is
obtained by taking $\W=z_\Omega(\H(B^{\mathrm c},\Omega))$ in the above
theorem and applying Proposition \ref{prop:reduce}.

\proof As a warm-up, let us prove that, if
$|\psi\rangle=a^\dagger(\xi')|0\rangle$, for some $\xi'\in\V$, then
(\ref{eq:locvac}) holds if and only if $\xi'\in(\overline{\mathrm{span}}_\C
\W)^\perp$ and $\parallel \xi'\parallel=1$.   Indeed, for all $\xi\in\W$,
\begin{eqnarray*}
\langle \psi|\Wf(\xi)|\psi\rangle &=&\langle 0|a(\xi')[\bbbone + a^\dagger(\xi)][\bbbone -a(\xi)]a^\dagger(\xi')|0\rangle \e^{-\frac12\parallel\xi\parallel^2}\\
&=&\langle0|\Wf(\xi)|0\rangle\parallel\xi'\parallel^2 - \langle 0|a(\xi')a^\dagger(\xi) a(\xi)a^\dagger(\xi')|0\rangle \e^{-\frac12\parallel\xi\parallel^2}\\
&=&\langle0|\Wf(\xi)|0\rangle\left[\parallel\xi'\parallel^2-
(\overline\xi'\cdot\xi) (\overline\xi\cdot\xi')\right].
\end{eqnarray*}
Supposing (\ref{eq:locvac}) holds, this clearly implies
$\parallel\xi'\parallel=1$ and $\xi'\in(\overline{\mathrm{span}}_\C
\W)^\perp$. The converse is equally obvious. This  proves the theorem for the
very particular case of states containing exactly one quantum.  Note that this
completely characterizes the states with exactly one field quantum that are
``localized''.

To prove part (i), we can now proceed as follows. Recall that
$$
\Wf(\xi)= \e^{-\frac12\parallel\xi\parallel^2}\e^{a^\dagger(\xi)}\e^{-a(\xi)}.
$$
Let $\xi\in\W$. Suppose $\psi=(\psi_0, \psi_1, \psi_2, \dots,
\psi_N, 0, 0,\dots)\in\eff^{\mathrm{fin},+}
((\overline{\mathrm{span}}_\C \W)^\perp)$. Then
$$
\langle\psi|\Wf(\xi)|\psi\rangle=\e^{-\frac12\parallel\xi\parallel^2}
\langle\psi, \psi \rangle=\langle0|\Wf(\xi)|0\rangle.
$$
Indeed, as a result of the fact that $\xi\in\W$ and
$\psi\in\eff^{\mathrm{fin},+}((\overline{\mathrm{span}}_\C
\W)^\perp)$, it follows that $a(\xi)\psi=0$ so that
$\e^{-a(\xi)}\psi=\psi$. From this one can conclude as follows.
For any $\psi=(\psi_0,\dots, \psi_n, \dots)\in\eff^+(
(\overline{\mathrm{span}}_\C \W)^\perp)$ and for any $N\in\N$, we
can write
$$
\psi = \psi_{<N} + \psi_{>N}
$$
where $\psi_{<N}=(\psi_0,\dots , \psi_N, 0, \dots, )$. Then, for any
$\epsilon>0$, there exists $N_\epsilon\in\N$ so that
$$
\langle \psi|\Wf(\xi)|\psi\rangle =
\langle\psi_{<N_\epsilon}|\Wf(\xi)|\psi_{<N_\epsilon}\rangle +
\mathcal{O}(\epsilon)=\langle0|\Wf(\xi)|0\rangle + \mathcal{O}(\epsilon),
$$
where the error term is uniform in $\xi$. Taking $\epsilon$ to $0$, the result
now follows.

In order to prove part (ii), I start with the following
preliminary computation. Let $N\in\N$ and consider $\psi=(\psi_0,
\psi_1, \psi_2, \dots, \psi_N, 0,
0,\dots)\in\eff^{\mathrm{fin},+}(\V)$ with $\psi_N\not=0$. We wish
to compute, for any $t\in\R$, for any $\xi\in \W$,
$$
\langle \psi | \Wf(t\xi) |\psi\rangle =\sum_{n,m=0}^N \langle \psi_n |
\Wf(t\xi) |\psi_m\rangle.
$$
We will first establish that
$$
\langle \psi | \Wf(t\xi) |\psi\rangle\e^{\frac12 t^2\parallel\xi\parallel^2}
$$
is a polynomial of degree at most $2N$ in $t$, for fixed $\xi$. For that
purpose, it is enough to notice that any term of the type
$$
\langle \psi_n | \Wf(t\xi) |\psi_m\rangle\e^{\frac12
t^2\parallel\xi\parallel^2}
$$
is a polynomial of degree at most $n+m$. This follows from
\begin{eqnarray*}
\langle \psi_n | \Wf(t\xi) |\psi_m\rangle\e^{\frac12 t^2\parallel\xi\parallel^2}&=&\langle \psi_n | \e^{a^\dagger(t\xi)} \e^{-a(t\xi)} |\psi_m\rangle\\
&=&\sum_{\ell_1=0}^{n} \sum_{\ell_2=0}^{m}\frac{1}{\ell_1!\ell_2!}\langle
\psi_n|(a^\dagger(t\xi))^{\ell_1}(-a(t\xi))^{\ell_2}|\psi_m\rangle
\end{eqnarray*}
It is clear that this is a polynomial of degree at most $n+m$. Also, the sum
can actually be restricted to those $\ell_1,\ell_2$ for which
$$
m-\ell_2=n-\ell_1.
$$

The term of degre $2N$ of the above polynomial is now  easily identified:
\begin{eqnarray*}
\langle \psi_N|\Wf(t\xi)|\psi_N\rangle  \e^{\frac12 \parallel t\xi\parallel^2}&=&\langle \psi_N| \e^{a^\dagger(t\xi)}\e^{-a(t\xi)}|\psi_N\rangle\\
&=&\sum_{\ell_1, \ell_2=0}^N \frac{1}{\ell_1!\ell_2!}\langle \psi_N|(a^\dagger(t\xi))^{\ell_1}(-a(t\xi))^{\ell_2}|\psi_N\rangle\\
&=&\frac{(-1)^N t^{2N}}{N! N!}\langle \psi_N|(a^\dagger(\xi))^N(a(\xi))^N |
\psi_N\rangle  + {\mathcal O}(t^{2N-1}).
\end{eqnarray*}
Suppose now (\ref{eq:locvac}) holds for $\psi$. Then this polynomial actually
has to be a constant, so, if $N\geq 1$,
$$
(a(\xi))^N|\psi_N\rangle=0
$$
for all $\xi\in\W$. Now let $\xi_1\dots\xi_N\in\W$ and consider the polynomial
$$
(a(t_1\xi_{1} + \dots + t_N \xi_{N}))^N|\psi_N\rangle=0
$$
in the variables $t_1, \dots, t_N\in\R$. Since each of its coefficients must
 vanish, we conclude that
$$
a(\xi_{1})a(\xi_{2})\dots a(\xi_{N})|\psi_N\rangle =0,
$$
for any choice of the $\xi_1\dots\xi_N\in\W$. Consequently, this is also true
for any choice of $\xi_1\dots\xi_N\in\mathrm{span}_\C \W$. Introduce now an
orthonormal basis $\eta_i$, $i\in\N$, of $\V$, with each
$\eta_i\in\mathrm{span}_\C \W$. Then, in view of the above,
$$
a(\eta_{\I _1})a(\eta_{\I _2})\dots a(\eta_{\I _N})|\psi_N\rangle
=0,
$$
for any choice $i_1\dots i_N\in\N$. It is then  clear that $\psi_N=0$. Since
by hypothesis $\psi_N\not=0$, it follows that $N=0$, so that $\psi$ belongs to
the zero-particle subspace $\eff_0^+(\V)$.
\endproof

\end{document}

%% file: chainnew.pstex_t
\begin{picture}(0,0)%
\includegraphics[height=1cm, keepaspectratio]{chainnew.pstex}%
\end{picture}%
\setlength{\unitlength}{3947sp}%
\begingroup\makeatletter\ifx\SetFigFont\undefined%
\gdef\SetFigFont#1#2#3#4#5{%
  \reset@font\fontsize{#1}{#2pt}%
  \fontfamily{#3}\fontseries{#4}\fontshape{#5}%
  \selectfont}%
\fi\endgroup%
\begin{picture}(6324,520)(5839,-2155)
\end{picture}